\documentclass[10pt,twocolumn,letterpaper]{article}

\usepackage{cvpr} 
\usepackage{times}
\usepackage{epsfig}
\usepackage{graphicx}
\usepackage{amsmath}
\usepackage{amssymb}
\usepackage{algorithm2e}
\usepackage{color}
\usepackage{authblk}

\usepackage{multirow}
\usepackage{graphicx}

\usepackage[breaklinks=true,bookmarks=false]{hyperref}

\cvprfinalcopy 

\def\cvprPaperID{****} 

\definecolor{dblue}{rgb}{0.0,0.0,0.5}
\definecolor{dgreen}{rgb}{0.0,0.5,0.0}
\definecolor{dred}{rgb}{0.5,0.0,0.0}

\DeclareMathOperator*{\argmax}{argmax}
\DeclareMathOperator*{\argmin}{argmin}

\renewcommand{\paragraph}[1]{\noindent\textbf{#1}\ }

\begin{document}

\title{Estimating Homogeneous Data-driven BRDF Parameters from a Reflectance Map under Known Natural Lighting}

\author
{
  Victoria L. Cooper  \quad  James C. Bieron \quad Pieter Peers\\
  College of William \& Mary\\
}

\maketitle

\def\wi{\omega_i}
\def\wo{\omega_o}
\def\fr{{\rho}}
\def\b{b}
\def\B{B}
\def\c{\zeta}
\def\w{w}
\def\I{I}
\def\L{L}
\def\P{P}
\def\LP{\log P}
\def\o{y}
\def\O{Y}
\def\p{p}
\def\n{n}
\def\N{\mathcal{N}}
\def\z{z}
\def\m{\gamma}
\def\U{U}
\def\S{S}
\def\V{V}
\def\cnt{c}

\begin{abstract}

In this paper we demonstrate robust estimation of the model parameters
of a fully-linear data-driven BRDF model from a reflectance map under
known natural lighting. To regularize the estimation of the model
parameters, we leverage the reflectance similarities within a material
class.  We approximate the space of homogeneous BRDFs using a Gaussian
mixture model, and assign a material class to each Gaussian in the
mixture model.  We formulate the estimation of the model parameters as
a non-linear maximum a-posteriori optimization, and introduce a linear
approximation that estimates a solution per material class from which
the best solution is selected.  We demonstrate the efficacy and
robustness of our method using the MERL BRDF database under a variety
of natural lighting conditions, and we provide a proof-of-concept
real-world experiment.

\end{abstract}

\section{Introduction}
\label{sec:intro}
Data-driven appearance models~\cite{Matusik:2003:DRM,Matusik:2003:EIB}
express the Bidirectional Reflectance Distribution Function (BRDF) of
a homogeneous material as a linear combination of a large set of
measured \emph{``basis''} BRDFs. The key assumption is that this large
set of basis BRDFs covers the full space of BRDFs, and any BRDF in
this space can be represented as convex combination of these basis
BRDFs, thereby inheriting all the intricate reflectance details
present in the measured basis BRDFs that can be difficult to model
with analytical BRDF models.  Compared to analytical BRDF models that
require an expensive and fragile non-linear optimization to estimate
the model parameters from reflectance measurements, data-driven BRDF
models, by virtue of its linear nature, only require a linear least
squares to estimate the model parameters. Recent advances have shown
great promise in reconstructing a data-driven BRDF from very few
measurements~\cite{Nielsen:2015:OMB,Xu:2016:MBS}.  However, these
methods rely on controlled directional or point lighting.  A key
problem in generalizing prior methods to natural lighting is that
these prior methods require a non-linear encoding (e.g., logarithmic)
to compress the dynamic range of the the basis BRDFs in order to
regularize the estimation of the model parameters. Such non-linear
encoding can only be undone after linear parameter estimation if the
observations consist of direct BRDF observations (i.e., a single view
and a single light direction per observation). In contrast,
observations under natural lighting are the result of an integration
of the BRDF times lighting over all directions, and only linear
transformations of the BRDF are transparent to this integration.

In this paper we aim to narrow the gap between inverse rendering with
data-driven BRDF models and analytical BRDF models under natural
lighting while retaining the robustness and simplicity of linear
parameter estimation for data-driven models.  We consider our work a
first exploration in this direction that demonstrates that robust
linear data-driven BRDF model parameter estimation under natural
lighting is feasible, rather than introducing a practical and/or
competitive method to current advanced inverse rendering methods that
use an analytical BRDF models as a basis.  To focus our exploration,
we will a-priori assume that the natural lighting is known and that we
have a full characterization of the material reflectance under this
lighting condition in the form of a reflectance
map~\cite{Rematas:2016:DRM}.

We desire to retain the advantages of a linear parameter estimation
process, and therefore avoid non-linear encoded basis BRDFs, and
directly estimate the data-driven BRDF model parameters from
unmodified basis BRDFs. To regularize the estimation of the model
parameters from a reflectance map under natural lighting, we leverage
the reflectance similarities between BRDFs in a material class.
Intuitively, we expect that it is easier to express the BRDF as a
combination of a small set of similar materials than from a large set
of BRDFs that span a larger spectrum of more varied materials.  We
therefore, first approximate the space of homogeneous BRDFs with a
Gaussian mixture model. Each normal distribution in the Gaussian
mixture model represents a material class, and we assign each basis
material to the class with the highest likelihood.  We formulate the
estimation of the model parameters as a maximum a-posteriori
optimization that maximizes the likelihood that the model parameters
explain the observations, as well as the likelihood that the model
belongs to the material class.  However, this formulation is highly
non-linear and difficult to minimize. We therefore exploit the
additional observation that in high dimensional spaces everything is
distant, and approximate the maximum a-posteriori optimization by an
efficient linear least squares approximation per material
class. Finally, we select the most likely provisional least squares
solution based on the maximum a-posteriori error.

We demonstrate the efficacy of our solution using the MERL BRDF
database under a variety of natural lighting conditions.  Furthermore,
we provide a proof-of-concept real-world experiment to demonstrate
that our results generalize beyond the ideal simulated experiments on
the MERL BRDF database.

\section{Related Work}
\label{sec:related}

We focus this discussion of prior work on the two key properties of
our method: reflectance modeling under natural lighting, and
appearance modeling with a data-driven reflectance model.  We refer to
the surveys of Dorsey~\etal~\cite{Dorsey:2007:DMM}, and Weinmann and
Klein~\cite{Weinmann:2015:AGR} for an in-depth general overview of
appearance modeling.

\paragraph{Reflectance Modeling under Natural Lighting}
A first subset of methods models surface reflectance from multiple
photographs under natural lighting. Oxholm and
Nishino~\cite{Oxholm:2016:SRE} model shape and homogeneous reflectance
from multiple photographs under known natural
lighting. Palma~\etal~\cite{Palma:2012:SMS},
Dong~\etal~\cite{Dong:2014:ARS}, and Zhou~\etal~\cite{Zhou:2016:SPS}
recover spatially-varying surface reflectance under \emph{unknown}
natural lighting from a dense sampling of multiple views or multiple
rotations of a subject with known shape.
Xia~\etal~\cite{Xia:2016:RSS} extended the method of
Dong~\etal~\cite{Dong:2014:ARS} to model spatially-varying reflectance
under unknown natural lighting \emph{and} unknown shape.  These model
all rely on non-linear reflectance models and estimation processes.
In contrast, we employ a linear data-driven BRDF model and rely on a
linear estimation process.

A second subset of methods models surface reflectance from just a
\emph{single} photograph of an object under natural lighting.  In
seminal work, Ramamoorthi and Hanrahan~\cite{Ramamoorthi:2001:SFI} lay
out a spherical harmonics framework for estimating general homogeneous
reflectance functions modeled by a spherical harmonics
expansion. Romeiro~\etal~\cite{Romeiro:2008:PR,Romeiro:2010:BR} model
the homogeneous surface reflectance using a bivariate data-driven
model from an object with known shape under known and unknown natural
lighting respectively. Similarly,
Lombardi~\etal~\cite{Lombardi:2016:RIR} also estimate natural lighting
and homogeneous surface reflectance modeled by the DSBRDF reflectance
mo\-del~\cite{Nishino:2011:DSB}.  Finally, Barron and
Malik~\cite{Barron:2015:SIR} recover shape, lighting and
spatially-varying albedo from a single photograph under unknown
natural lighting. However, Barron and Malik only consider diffuse
reflectance.  Our method espouses the same overall goal as this second
subset of methods. A reflectance map can potentially be obtained from
a single observations of a convex object of known shape (e.g., sphere)
or using the deep learning method of
Rematas~\etal~\cite{Rematas:2016:DRM}. However, we explicitely desire
to recover a data-driven model~\cite{Matusik:2003:DRM} based on
real-world measured reflectance.

A third subclass of methods relies on deep learning to infer
reflectance properties under unknown natural lighting from a single
image. Li~\etal~\cite{Li:2017:MSA} and Ye~\etal~\cite{Ye:2018:SPS}
estimate the parameters of an analytical BRDF
model~\cite{Ward:1992:MMA} for a spatially-varying material.  Both
Li~\etal and Ye~\etal focus on augmenting the training data with
unlabeled photographs in order to reduce the number of required
labeled training data (i.e., measured
SVBRDFs). Li~\etal~\cite{Li:2018:MMS} present a network structure and
a novel post-processing step based on conditional random fields to
estimate spatially-varying reflectance parameters for an analytical
micro-facet BRDF model~\cite{Walter:2007:MMR}. Finally,
Li~\etal~\cite{Li:2018:LRS} propose a cascading network structure to
iteratively estimate and refine the shape and spatially-varying
surface reflectance.  All of the above methods express the surface
reflectance using an analytical BRDF model. In contrast, we express
the surface reflectance using a more expressive data-driven model,
albeit limited to a homogeneous material and under \emph{known}
natural lighting.

\paragraph{Data-driven Reflectance Model}
In seminal work, Matusik~\etal~\cite{Matusik:2003:DRM} presented a
data-driven BRDF model that expresses the surface reflectance as a
weighted combination of a large set of measured BRDFs.  To handle the
large dynamic range between the specular peaks and the diffuse
reflectance, a log-encoding is first applied to the measured basis
BRDFs.  Matusik~\etal propose two models: a PCA based $45$D linear
model, and non-linear, charting based, $15$D model.  In follow up
work, Matusik~\etal~\cite{Matusik:2003:EIB} use the linear PCA model
and show that $800$ well selected and controlled view-light direction
pairs are sufficient for estimating the BRDF.
Nielsen~\etal~\cite{Nielsen:2015:OMB} show that by adding a Tikhonov
regularization to the estimation of a log-relative encoded linear
data-driven model, a good BRDF estimate can be obtained from less than
$20$ optimized and controlled view-light direction pairs, and for $5$
photographs of a sphere lit by optimized directional light sources.
Xu~\etal~\cite{Xu:2016:MBS} build on the method of Nielsen~\etal, and
show that with an improved error metric, a log-relative encoded linear
data-driven model can be recovered from just $2$ near-field
observations (photographs) under controlled directional lighting. All
of the above methods estimate a data-driven BRDF from observations
under directional lighting, and regularize the estimation using a
non-linear encoding of the measured BRDFs.  In contrast, our method
uses a fully linear model and reconstructs the data-driven BRDF model
from a reflectance map under uncontrolled known natural lighting.

\section{Overview}
\label{sec:overview}

\paragraph{Data-driven BRDF}
The reflectance behavior of a homogeneous material is described by
the bidirectional reflectance distribution function (BRDF) $\fr(\wi,
\wo)$: a $4D$ function defined as the ratio of incident irradiance
for an incident direction $\wi$ over the outgoing radiance for an
outgoing direction $\wo$.

In this paper, we follow the data-driven BRDF model of
Matusik~\etal~\cite{Matusik:2003:EIB} that characterizes the BRDF
$\fr$ as a linear combination of a large set of $n$ measured materials
$\b_i, i \in [1, n]$.  The underlying idea is that the set of measured
BRDFs spans the space of BRDFs, and any material's BRDF should lie in
this space:
\begin{equation}
  \fr = \B \w,
\label{eq:datadriven}
\end{equation}
where we stack the BRDF $\fr$ and basis BRDFs $\b_i$ in a vector of
length $\p$, and form the matrix $\B$ by stacking each basis vector in
a column: $\B = [ \b_1 , ..., \b_\n ]$. The model parameters are
stacked in a vector $\w$ of $\n$ scalar weights.  We directly use the
BRDF parameterization of the MERL BRDF
database~\cite{Matusik:2003:DRM}, and $\p = 90 \times 90 \times 180$.
Furthermore, similar as in Nielsen~\etal~\cite{Nielsen:2015:OMB}, we
consider each color channel of the $100$ MERL BRDFs as a basis BRDF,
and thus $\n = 300$.

Due to the large dynamic range between specular peaks versus diffuse
reflectance, prior
work~\cite{Matusik:2003:DRM,Nielsen:2015:OMB,Xu:2016:MBS} has applied
a non-linear compression function $\c$ to make the estimation of $\w$
less sensitive to errors on the (large) specular peaks:
\begin{equation}
  \fr' = \B' \w',
\label{eq:compress}
\end{equation}
where $B' = [ \c(\b_1), ..., \c(\b_\n) ]$. An expansion $\c^{-1}$ is
applied to the compressed BRDF $\fr'$ after computation of the
weights. A common compression function is the logarithmic function, in
which case~\autoref{eq:compress} becomes a homomorphic
factorization.

\paragraph{Natural Lighting}
Prior work relied on point sample measurements of $\fr$ for a set of
incoming-outgoing direction pairs to estimate the weights $\w$. In
contrast, in this paper we aim to estimate the weights $\w$ from an
observation under natural lighting. Assuming the lighting $\L$ is
distant (i.e., it only depends on the incident direction $\wi =
(\phi_i, \theta_i)$), and ignoring interreflections, we can formulate
the observed radiance $\o$ as:
\begin{equation}
  \o(\wo) = \int_\Omega \fr(\wi, \wo) \cos(\theta_i) \L(\wi) d\wi,
\label{eq:render}
\end{equation}
where $\cos(\theta_i)$ is the foreshortening, and $\Omega$ is the
upper hemisphere of incident directions. Due to linearity of light
transport, we can express~\autoref{eq:render} in terms of
corresponding basis observations $\o$:
\begin{equation}
  \o = \O \w,
\label{eq:render2}
\end{equation}
where the weights $\w$ are the same as in~\autoref{eq:datadriven}, and
thus can be used to reconstruct $\fr$. The basis images $\O = [\o_0,
  ..., \o_n]$ are the observations of the measured basis BRDFs $\b_i$
under the same conditions:
\begin{equation}
  \o_i = \int_\Omega \b_i(\wi, \wo) \cos(\theta_i) \L(\wi) d\wi.
\label{eq:renderbasis}
\end{equation}

\paragraph{Problem Statement}
As noted before, the dynamic range compression function $\c$ is
essential in obtaining good data-driven BRDF reconstructions, even in
the case of a very dense point sampling of light and view
directions~\cite{Bagher:2016:ANP}. However, this compression function
cannot be used when linearly estimating the weights $\w$ from
observations under natural lighting. This can be seen by
inserting~\autoref{eq:compress} in~\autoref{eq:renderbasis}:
\begin{equation}
 \c(\o_i) \neq \o^\c_i = \int_\Omega \c(\b_i(\wi, \wo)) \cos(\theta_i) \L(\wi) d\wi.
\end{equation}
In other words, the non-linear compression of the observation is not
equivalent to the observation under natural lighting of the
non-linearly compressed BRDFs. While not a problem for the basis BRDFs
$\b_i$, since we can generate the corresponding images $\o^\c_i$ with
any rendering system directly from the non-linear encoded basis BRDFs
$\c(b_i)$, it is a problem for $\fr$, because we can only observe $\o$
the resulting radiance of $\fr$ under natural lighting, not the
reflected radiance of its non-linear compressed form $\c(\fr)$, and
hence we do not have access to $\o^\c$.  Consequently, the key problem
we aim to address in this paper is to find the data-driven weights
$\w$ from the observation $\o$ without relying on a non-linear
compression function $\c$ and/or a non-linear optimization procedure
for estimating the weights $\w$.

\paragraph{Maximum a-posteriori Optimization}
Formally, our goal is to find the most likely weights $\w$, relying on
a linear estimation process, such that the conditional probability of
the reconstructed data-driven homogeneous BRDF $\fr$ is maximized
given a reflectance map $\o$ under known natural lighting $\L$:
\begin{equation}
  \argmax_\w \P(\fr | \o).
\end{equation}
We will assume that the observations are in the form of a high dynamic
range reflectance map (i.e., a full characterization of the
reflectance radiance of a homogeneous BRDF for a fixed lighting
condition). In the remainder of this paper, we will assume that the
reflectance map is provided in the form of a visualization of a sphere
under the target illumination.

Using Bayes' theorem, we can formulate the maximum a-posteriori (MAP)
estimation of $\w$ as:
\begin{equation}
  \argmax_\w \frac{\P(\o | \fr) \P(\fr)}{\P(\o)}.
\end{equation}
Rewriting in terms of the log-likelihood, and noting that $\P(\o)$ is
constant (i.e., the observation is given), we obtain:
\begin{equation}
  \argmin_\w \left( \LP(\o | \fr) + \LP(\fr) \right).
\label{eq:map}
\end{equation}

In order to solve this minimization problem, we need a model of the
likelihood of the BRDF estimation $\fr$ (\autoref{sec:clustering}),
and a model for the conditional probability of the observation $\o$
given the estimated BRDF $\fr$, and an efficient linear strategy for
solving this minimization (\autoref{sec:reconstruction}).

\section{BRDF Likelihood Modeling}
\label{sec:clustering}

\paragraph{Gaussian Mixture Model}
We propose to model the likelihood of BRDFs by a Gaussian mixture
model (GMM):
\begin{equation}
  \P(\fr) = \sum^k_{j=1} \pi_j \N(\fr | \mu_j, \Sigma_j),
\label{eq:gmm}
\end{equation}
where $\pi_j$ are the mixing coefficients of the $j$-th normal
distribution $\N$ with mean $\mu_j$ and covariance matrix $\Sigma_j$.

\paragraph{Expectation-Maximization}
An effective method for computing the parameters $\Theta = (\pi, \mu,
\Sigma)$ is the Expectation Maximization algorithm using the MERL
BRDFs $b_i$ as observations. For this we define a latent variable
$\m_j(\b_i)$ that indicates the likelihood of the $j$-th Gaussian
given a MERL BRDF $\b_i$:
\begin{eqnarray}
  \m_j(\b_i) & = & \P(j | \b_i), \\
            & = & \frac{\P(j) \P(\b_i | j)}{\P(\b_i)}, \\
            & = & \frac{\pi_j \N(\b_i | \mu_j, \Sigma_j)}{\sum^k_{j=1} \pi_j \N(\b_i | \mu_j, \Sigma_j)}.
\label{eq:E}
\end{eqnarray}
Expectation minimization iterates between estimating the latent
variable $\m_j(\b_i)$ (E-step, \autoref{eq:E}), and the model
parameters (M-step):
\begin{eqnarray}
  \pi_j & = & \frac{1}{n} \sum^n_i \m_j(\b_i), \\
  \mu_j & = & \frac{\sum^n_i \m_j(\b_i) \b_i}{\pi_j}, \\
  \Sigma_j & = & \frac{\sum^n_i \m_j(\b_i) (\b_i - \mu_j)(\b_i - \mu_j)^T}{\pi_j}.
\label{eq:M}
\end{eqnarray}
We iterate until the log-likelihood over the MERL BRDFs converges:
\begin{equation}
  \LP(B | \Theta) = \sum^n_i \log \sum^k_j \pi_j \N(\fr | \mu_j, \Sigma_j).
\end{equation}

To bootstrap the EM algorithm, we perform a standard k-mean
clustering, and initialize $\pi_j$ as the ratio of assigned BRDFs to
the $j$-th cluster over the total number of MERL BRDFs (i.e., $\n$).
    
\paragraph{Curse of Dimensionality}
A practical problem is that the number of observations $\n$ is
significantly lower than the dimensionality of the space (i.e.,
$\p$).We therefore apply a singular value decomposition (SVD) to
express the observations in a $\n$ dimensional space $\U$:
\begin{equation}
  \B = \U \S \V^T.
\end{equation}
However, this is still a $300$ dimensional space. A key issue is that
even for a moderate number of dimensions any distance is very large,
and thus the distance to the means $\mu_j$ are large
too. Consequently, the likelihood of each Gaussian mixture
(\autoref{eq:E}) will always be very low and it can potentially cause
numerical instabilities.  To resolve this issue, we perform
expectation maximization in a reduced space, and only keep the
coefficients belonging to the $N$ largest singular values. In other
words, we perform expectation maximization (i.e., soft clustering) on
a projection to an $N$ dimensional subspace, and approximate the
likelihood: $\P(\fr) \approx \P(\hat{\U}^T \fr)$, where $\hat{\U}$ is
the $N$ dimensional basis (i.e., the first $N$ vectors in $\U$).

\begin{figure}[t!]
\begin{center}
\includegraphics[width=0.45\textwidth]{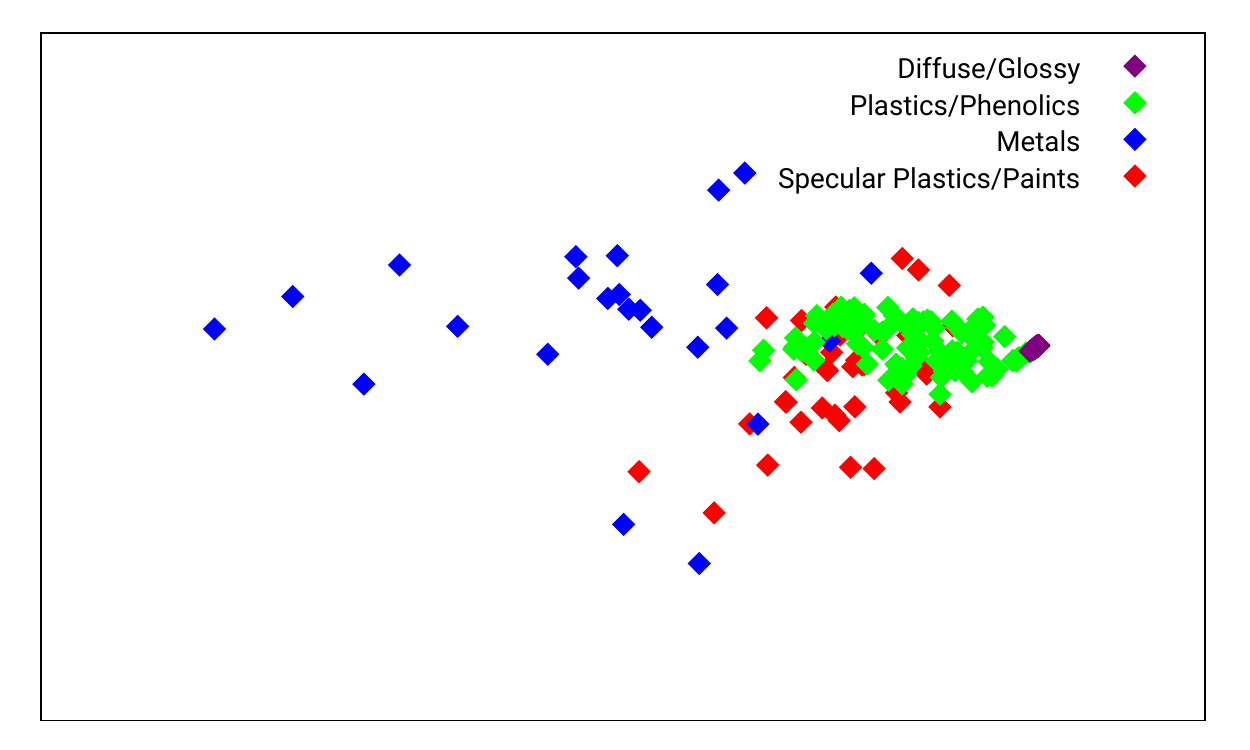}
\caption{$2$D multi-dimensional scaling of the projected MERL BRDFs
  $\hat{\U}^T \B$ and a color-coding of the respective material classes
  derived from the $4$D approximation of the BRDF likelihood modeled
  by a Gaussian mixture model.}
\label{fig:plot}
\end{center}
\end{figure}

\paragraph{Discussion}
We found that $N=4$ offers a good balance between accuracy and
numerical stability.  A second parameter that needs to be set is the
number of Gaussian mixtures $K$. If the number of Gaussians is too
low, then $P(\hat{\U}^T \fr)$ only offers a coarse
approximation. However, we also found that for increasing number of
$K$, the algorithm tends to subdivide the same Gaussian distribution,
essentially overfitting to 'special case' BRDFs (such as \emph{Steel}
which exhibits acquisition artifacts).  In practice we found that
$K=4$ offers a good approximation that nicely categorizes the
materials in four recognizable distinct material classes:
\emph{``diffuse and glossy''} materials ($137$ materials),
\emph{``plastics/phenolics''} ($99$ materials), \emph{``metals''}
($24$), and \emph{``specular plastics/paints''} ($40$ materials); we
determine membership to a material class by assigning the material to
the material class with the maximum $\m_j(b_i)$ likelihood.
\autoref{fig:plot} shows a plot of a $2$D multi-dimensional scaling of
the $4$D projected coordinates of the MERL BRDFs, as well as a
color-coding to indicate for which material class the material has the
highest affinity.  Note that even though the diffuse-like material
class contains $137$ materials, the multi-dimensional scaling places
them all close together.  Please refer to the supplemental material
for an exhaustive list of which material belongs to which material
class.

\section{Data-driven Model Estimation}
\label{sec:reconstruction}

\paragraph{MAP Estimation}
We express the likelihood of the observation given an estimate of the
BRDF as:
\begin{equation}
  \P(\o | \fr) = \N(\O \w - \o | \mu, \sigma),
\label{eq:dataterm}
\end{equation}
where $\mu$ and $\Sigma$ is the expected mean error and standard
deviation on the reconstructions, and $\O\w$ is the rendering of the
estimated BRDF under the target natural lighting.  We assume that the
mean error is close to zero ($\mu = 0$), and $\sigma$ is proportional
to the expected measurement error (e.g., camera noise).

Given the likelihood $\P(\hat{U}^T \fr)$ expressed by the Gaussian
mixture model (\autoref{eq:gmm}), we can then formulate the MAP
estimation (\autoref{eq:map}) as:
\begin{equation}
  \argmin_\w \left( \frac{||\O\w - \o||^2}{\sigma^2} + \log \sum_j \pi_j \N( \hat{\U}^T \B \w | \mu_j, \Sigma_j) \right).
\label{eq:nonlinloss}
\end{equation}
The first term is the data term that indicates how well (a
visualization of) the BRDF $\fr = \B\w$ can explain the observation
$\o$, and the second term indicates how plausible the reconstructed
BRDF $\fr$ (projected in the $4$ dimensional space $\hat{\U}$) is.

However, directly solving for the BRDF weights $\w$ using
\autoref{eq:nonlinloss} is not practical because of two key issues:
\begin{enumerate}
  \item \emph{Non-linear:} \autoref{eq:nonlinloss} is highly
    non-linear and difficult to optimize due to the sum of the
    log-likelihoods in the second term.
  \item \emph{Gaussian Mixture Model Accuracy for $\P(\fr) \approx
    \P(\hat{\U}^T \fr)$:} We approximated the likelihood of the BRDF
    by a $4$ dimensional Gaussian mixture model.  This reduction in
    dimensionality was necessary due to the curse of
    dimensionality. However, it also implicitly assumes that the BRDF
    lies not too far from the space of plausible BRDFs.  Since the
    likelihood is only determined based on $4$ dimensions (and thus
    only regularizes these four), the other $296$ dimensions can be
    set to any value (including unreasonable values that result in an
    implausible BRDF).
\end{enumerate}

\paragraph{Linear MAP Approximation}
To alleviate the above two practical issues, we exploit the
observation that the likelihood of a basis BRDF $\b_i$ belonging to a
material class $m$ is for most basis BRDFs equivalent to an indicator
function:
\begin{equation}
  \m_j(\b_i) \approx \delta_{i,m}.
\end{equation}
This implies that the overlap between the Gaussians in the Gaussian
mixture model is limited.  Armed with this observation, we therefore
propose to compute a candidate BRDF for each material class $j \in [1,
  k]$:
\begin{equation}
  \argmin_{\w^{(j)}} \left( \LP(\o | \fr, j) + \LP(\fr | j) \right).
\end{equation}
Given the set of candidate solutions $\w' = \{ \w^{(1)}, .., \w^{(k)}
\}$, we then pick the best candidate that best reconstructs the BRDF. 

\paragraph{Per-Material Class Linear Data Term}
We define the data-term similarly as in the general non-linear case,
except that we only use the basis BRDFs that belong to the same
material class:
\begin{eqnarray}
  & & \LP(\o | \fr, j ) = || \O^{(j)} \w^{(j)} - \o ||^2, 
\label{eq:matdataterm}
\end{eqnarray}
where $\O^{(j)}$ is the set of observations that correspond to the
basis BRDFs assigned to the $j$-th material class (i.e., the materials
$\b_i$ for which $\m_j(\b_i)$ is maximal).

\paragraph{Per-Material Class Linear Likelihood Term}
We express the per-material class likelihood by a single Gaussian
model.  We directly compute this probability on the BRDF weights
$\w^{(j)}$:
\begin{equation}
  \P(\fr | j) = \N(\w^{(j)}, \mu'_j, \Sigma'_j),
\label{eq:biasterm}
\end{equation}
where: $\mu'_j = (\frac{1}{\cnt_j}, ..., \frac{1}{\cnt_j})^T$, and
$\cnt_j$ is the number of basis BRDFs in the $j$-th material class.
Note that $\O^{(j)} \mu'_j$ is equivalent to the mean BRDF of the
material class, and $\mu'_j$ the corresponding coordinate in the j-th
BRDF subspace.

\paragraph{Linear Least Squares Estimation}
Both~\autoref{eq:matdataterm} and (the log likelihood of)
\autoref{eq:biasterm} are quadratic terms that define a linear system
in terms of $\w$ that can be solved using a regular linear least
squares.  However, both terms can have a vastly different magnitude.
The magnitude of the data-term depends on the error on the rendered
image of the estimated BRDF.  This image error depends on the
resolution, the overall intensity of the lighting, and the
reflectivity of the material. Similarly, the magnitude of the
likelihood term depends on the number of basis BRDFs per material
class.  We therefore add a balancing term:
\begin{equation}
  \lambda_j = \frac{\lambda ||\o||^2}{\cnt_j},
\label{eq:lambda}
\end{equation}
where $||\o||^2$ is the total squared pixel intensities in the
observation.  We expect that the overall intensity of the observation
is directly proportional to the lighting intensity and reflectivity of
the BRDF, and hence the overall scale of the image error. $\lambda$ is
a user set constant that depends on the qualities of the lighting.  An
ill-conditioned lighting condition requires a larger $\lambda$ value
(e.g., a low frequency lighting environment is ill-conditioned for
estimating specular properties~\cite{Ramamoorthi:2001:SFI}).  In
practice we found that $\lambda = 0.5$ works well for many lighting
environments, and forms a good starting point for fine-tuning
$\lambda$.

The final linear least squares is:
\begin{eqnarray}
  & & \argmin_{\w^{(j)}} \left( ||\O^{(j)} \w^{(j)} - \o||^2 + \lambda_j \frac{|| \w^{(j)} - \mu'_j ||^2}{\Sigma_j^2} \right).
\label{eq:lls}
\end{eqnarray}

\paragraph{Selection}
Ideally, we would like to select the best candidate solution from
$\w'$ by evaluating~\autoref{eq:nonlinloss}.  However, by a-priori
assuming that a BRDF belongs to a material class $j$, it is possible
that there is a significant mismatch between the target material and
the material class. For example, attempting to model a mirror-like
specular material using the diffuse material class is unlikely to
produce a satisfactory result.  Consequently, we cannot simply rely on
the likelihood $\P(\hat{U}^T \fr)$ based on the 4 dimensional Gaussian
mixture model to select the best solution from $\w'$ (i.e., the other
$296$ dimensions can be arbitrarily wrong). We will therefore further
exploit the observation of the limited overlap of the Gaussians in the
mixture model, and approximate the solution per material class by
enforcing that it lies in the convex hull of the subspace spanned by
the BRDFs assigned to the material class, and only rely
on~\autoref{eq:dataterm} to pick the best candidate from $\w'$. We
ignore the standard deviation (i.e., $\sigma = 1$) in
\autoref{eq:dataterm} as it only acts as a scale (in the
log-likelihood) that does not affect the selection of the best
reconstruction (i.e., minimum log-likelihood).

\paragraph{Color}
Our discussion until now only considered mono\-chro\-me BRDFs; we used
all color channels from the MERL BRDFs as separate basis BRDFs. A
straightforward strategy for estimating a non-monochrome BRDF with
three color channels, would be to execute the estimation separately
for each color channel, and combine the three reconstructed
mo\-no\-chro\-me BRDF into a single RGB BRDF.  However, it is possible
that a solution from a different material classes $j$ is selected for
each of the three color channels. Because the set of basis BRDFs for
each material class are disjunct, there can be slight differences in
the constructed BRDF shape for each color channel, which in turn can
result in color artifacts in the combined BRDF.  We circumvent this
potential problem by combining the three color channels after
obtaining the candidate BRDFs, and performing the selection on the RGB
BRDF instead of each color channel separately.  Hence, each color
channel will be reconstructed with the same set of basis BRDFs.

\begin{figure*}[ht!]
\begin{center}
{\setlength{\tabcolsep}{0mm}\small
\def\widthA{0.162\linewidth}

\begin{tabular}{ p{0.35cm} c c| c c| c c }
 & \multicolumn{2}{c|}{} & \multicolumn{2}{c|}{Reconstructed under} & \multicolumn{2}{c}{Reconstructed under} \\
 & \multicolumn{2}{c|}{Reference} & \multicolumn{2}{c|}{Eucalyptus Grove} & \multicolumn{2}{c}{Galileo} \\
{\rotatebox[origin=c]{90}{Spec. Orange Phenolic}}& 
\raisebox{-0.5\height}{\includegraphics[width=\widthA]{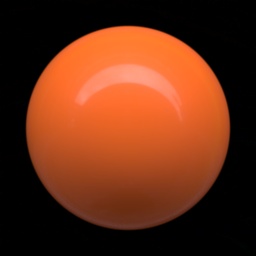}}&
\raisebox{-0.5\height}{\includegraphics[width=\widthA]{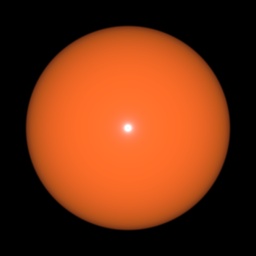}}&
\raisebox{-0.5\height}{\includegraphics[width=\widthA]{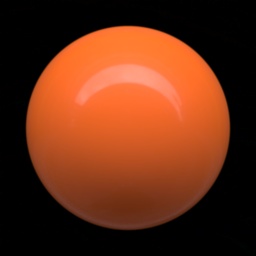}}&
\raisebox{-0.5\height}{\includegraphics[width=\widthA]{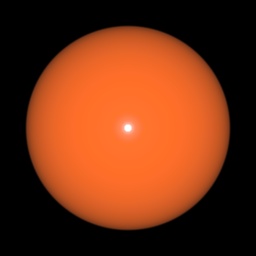}}&
\raisebox{-0.5\height}{\includegraphics[width=\widthA]{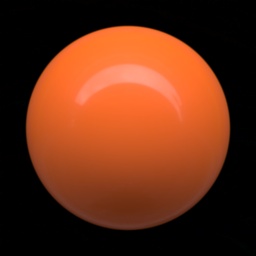}}&
\raisebox{-0.5\height}{\includegraphics[width=\widthA]{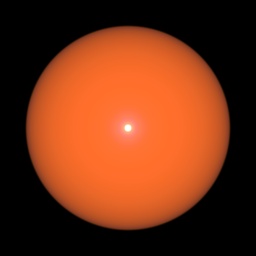}}\\

{\rotatebox[origin=c]{90}{Colonial Maple}}& 
\raisebox{-0.5\height}{\includegraphics[width=\widthA]{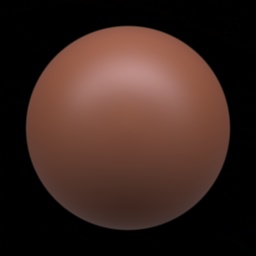}}&
\raisebox{-0.5\height}{\includegraphics[width=\widthA]{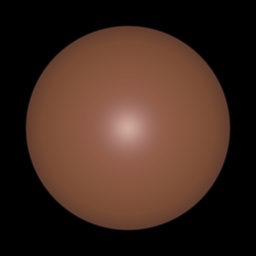}}&
\raisebox{-0.5\height}{\includegraphics[width=\widthA]{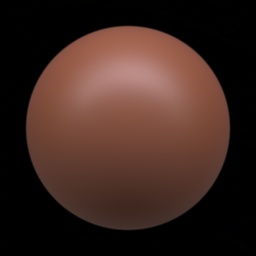}}&
\raisebox{-0.5\height}{\includegraphics[width=\widthA]{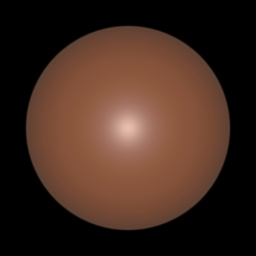}}&
\raisebox{-0.5\height}{\includegraphics[width=\widthA]{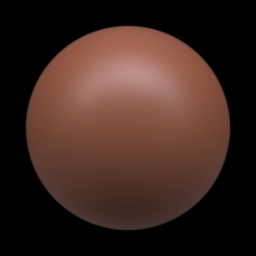}}&
\raisebox{-0.5\height}{\includegraphics[width=\widthA]{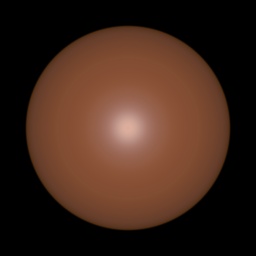}}\\

{\rotatebox[origin=c]{90}{Green Latex}}& 
\raisebox{-0.5\height}{\includegraphics[width=\widthA]{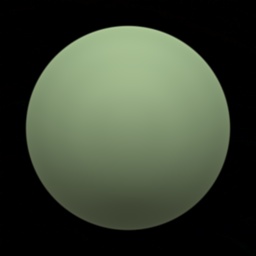}}&
\raisebox{-0.5\height}{\includegraphics[width=\widthA]{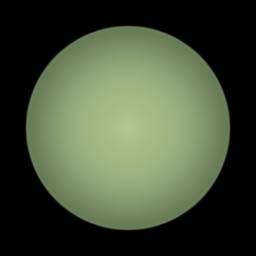}}&
\raisebox{-0.5\height}{\includegraphics[width=\widthA]{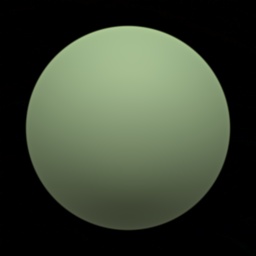}}&
\raisebox{-0.5\height}{\includegraphics[width=\widthA]{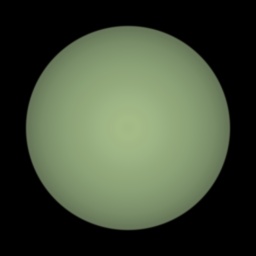}}&
\raisebox{-0.5\height}{\includegraphics[width=\widthA]{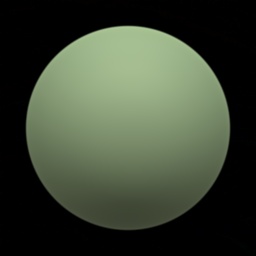}}&
\raisebox{-0.5\height}{\includegraphics[width=\widthA]{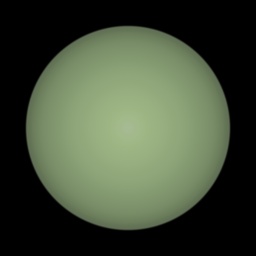}}\\

{\rotatebox[origin=c]{90}{Steel}}& 
\raisebox{-0.5\height}{\includegraphics[width=\widthA]{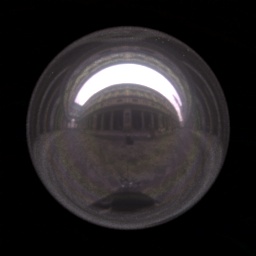}}&
\raisebox{-0.5\height}{\includegraphics[width=\widthA]{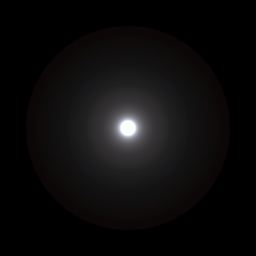}}&
\raisebox{-0.5\height}{\includegraphics[width=\widthA]{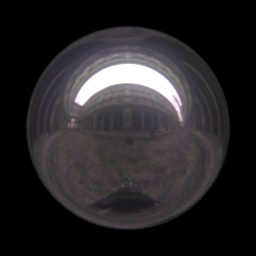}}&
\raisebox{-0.5\height}{\includegraphics[width=\widthA]{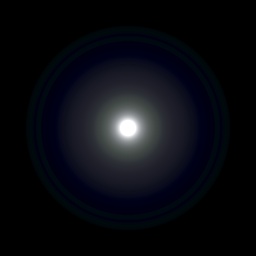}}&
\raisebox{-0.5\height}{\includegraphics[width=\widthA]{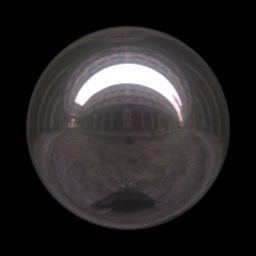}}&
\raisebox{-0.5\height}{\includegraphics[width=\widthA]{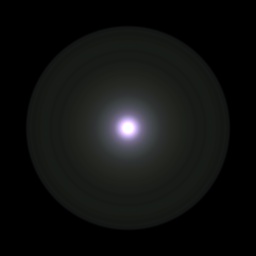}}\\

{\rotatebox[origin=c]{90}{Green Metal. Paint 2}}& 
\raisebox{-0.5\height}{\includegraphics[width=\widthA]{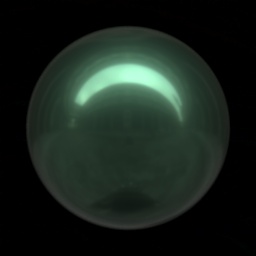}}&
\raisebox{-0.5\height}{\includegraphics[width=\widthA]{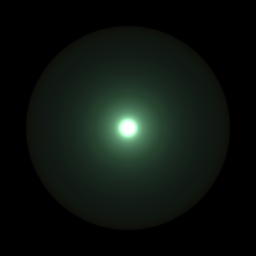}}&
\raisebox{-0.5\height}{\includegraphics[width=\widthA]{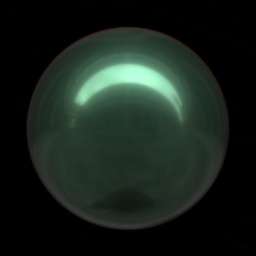}}&
\raisebox{-0.5\height}{\includegraphics[width=\widthA]{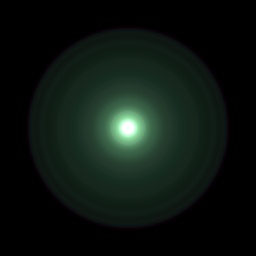}}&
\raisebox{-0.5\height}{\includegraphics[width=\widthA]{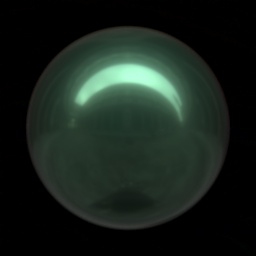}}&
\raisebox{-0.5\height}{\includegraphics[width=\widthA]{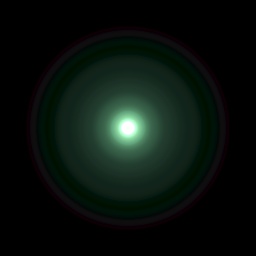}}\\

{\rotatebox[origin=c]{90}{Green Metal. Paint}}& 
\raisebox{-0.5\height}{\includegraphics[width=\widthA]{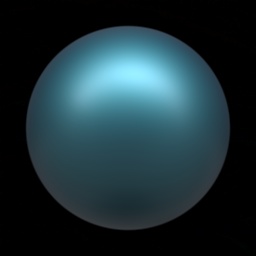}}&
\raisebox{-0.5\height}{\includegraphics[width=\widthA]{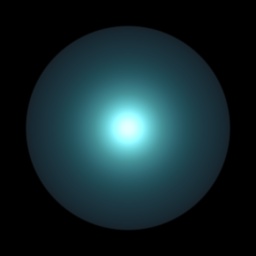}}&
\raisebox{-0.5\height}{\includegraphics[width=\widthA]{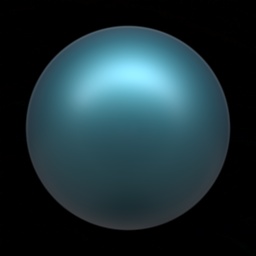}}&
\raisebox{-0.5\height}{\includegraphics[width=\widthA]{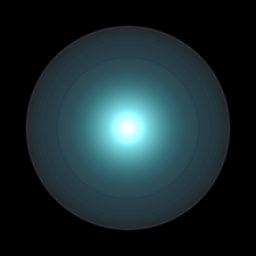}}&
\raisebox{-0.5\height}{\includegraphics[width=\widthA]{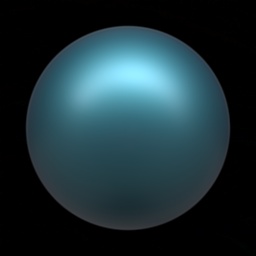}}&
\raisebox{-0.5\height}{\includegraphics[width=\widthA]{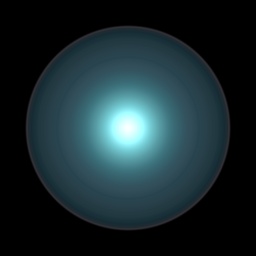}}\\

{\rotatebox[origin=c]{90}{Yellow Matte Plastic}}& 
\raisebox{-0.5\height}{\includegraphics[width=\widthA]{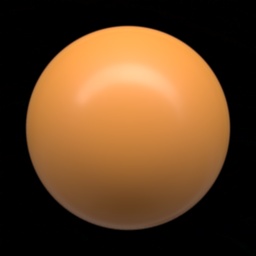}}&
\raisebox{-0.5\height}{\includegraphics[width=\widthA]{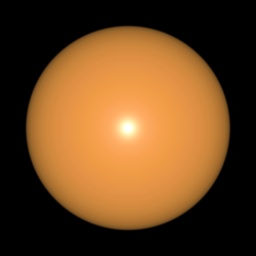}}&
\raisebox{-0.5\height}{\includegraphics[width=\widthA]{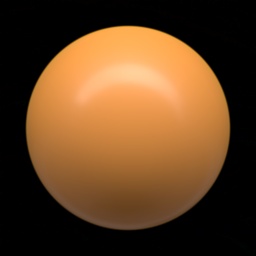}}&
\raisebox{-0.5\height}{\includegraphics[width=\widthA]{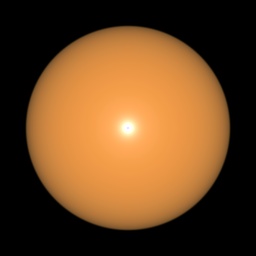}}&
\raisebox{-0.5\height}{\includegraphics[width=\widthA]{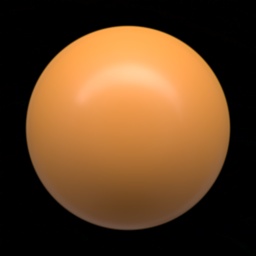}}&
\raisebox{-0.5\height}{\includegraphics[width=\widthA]{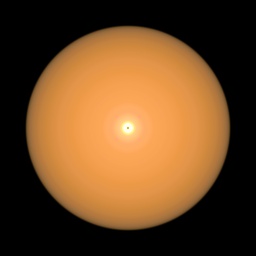}}\\

\end{tabular}
}
\caption{Data-driven BRDF reconstructions from a reflectance map under
  the \emph{Eucalyptus Grove} and the \emph{Galileo's Tomb} light
  probe. We visualize the reference and reconstructed BRDFs under the
  \emph{Uffizi Gallery} light probe and a directional light.}
\label{fig:fullpage}
\end{center}
\end{figure*}

\paragraph{Algorithm Summary}
In summary, given a reflectance map $\o$ under known natural lighting
$\L$, and given a user provided balance parameter $\lambda$, we
compute the data-driven BRDF $\fr = \B \w$ as:
\begin{enumerate}
  \item We precompute the Gaussian mixture model using the EM
    algorithm detailed in~\autoref{sec:clustering}. Note, this
    precomputation only needs to happen once for the MERL BRDF
    database, and is independent of the lighting.
  \item We precompute $\O$ by rendering a sphere with each basis BRDF
    $\b_i$ under the natural lighting (\autoref{eq:renderbasis}). This
    precomputation needs to happen for every lighting condition.
  \item We compute the candidate solutions $\w'_{\{r,g,b\}}$ for each
    material class by solving the linear least squares
    in~\autoref{eq:lls} per color channel.
  \item We combine the monochrome BRDFs to a 3-channel BRDF: $\w' = \{
    (\w'_{r,1}, \w'_{g,1}, \w'_{b,1}), ..., (\w'_{r,k}, \w'_{g,k},
    \w'_{b,k}) \}$.
  \item Finally, we select the candidate solution from $\w'$ that
    minimizes \autoref{eq:dataterm}.
\end{enumerate}

\section{Results}
\label{sec:results}

\begin{figure*}[ht!]
\begin{center}
{\setlength{\tabcolsep}{0mm}\small
\def\widthA{0.135\linewidth}
\begin{tabular}{ p{0.35cm} c| c| c c c c }
 &           & Linear Least & \multicolumn{4}{c}{Material Class}\\
 & Reference & Squares & Diffuse & Plastics/Phenolics & Metals & Spec. Plastics/Paints \\

\multirow{3}{*}{\rotatebox[origin=c]{90}{Dark Blue Paint\hspace{0.7cm}}}& 
\raisebox{-0.5\height}{\includegraphics[width=\widthA]{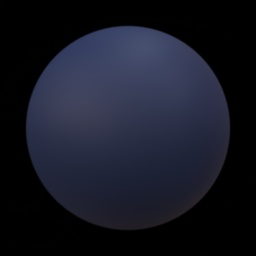}}&
\raisebox{-0.5\height}{\includegraphics[width=\widthA]{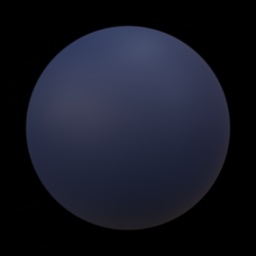}}&
\raisebox{-0.5\height}{\includegraphics[width=\widthA]{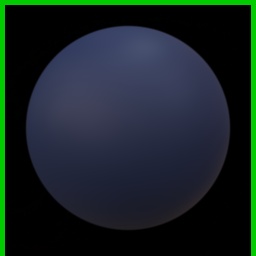}}&
\raisebox{-0.5\height}{\includegraphics[width=\widthA]{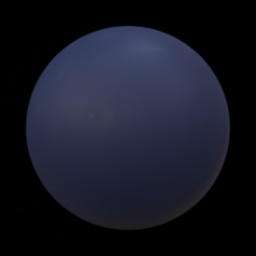}}&
\raisebox{-0.5\height}{\includegraphics[width=\widthA]{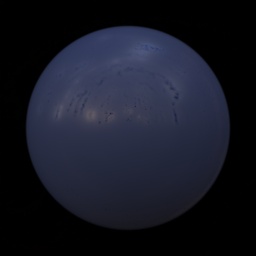}}&
\raisebox{-0.5\height}{\includegraphics[width=\widthA]{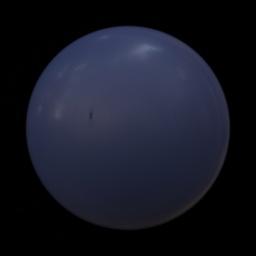}}\\
&
\raisebox{-0.5\height}{\includegraphics[width=\widthA]{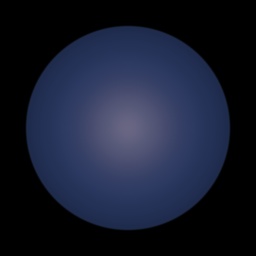}}&
\raisebox{-0.5\height}{\includegraphics[width=\widthA]{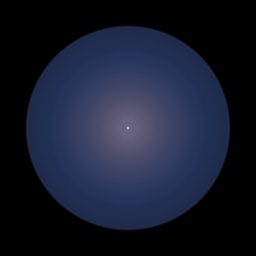}}&
\raisebox{-0.5\height}{\includegraphics[width=\widthA]{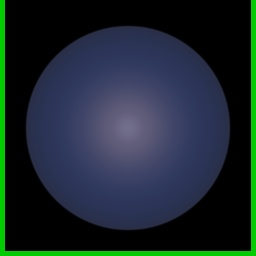}}&
\raisebox{-0.5\height}{\includegraphics[width=\widthA]{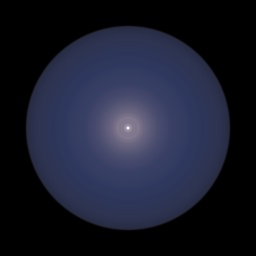}}&
\raisebox{-0.5\height}{\includegraphics[width=\widthA]{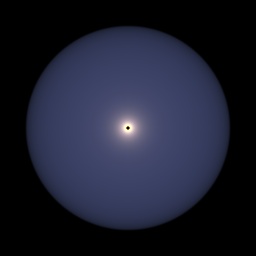}}&
\raisebox{-0.5\height}{\includegraphics[width=\widthA]{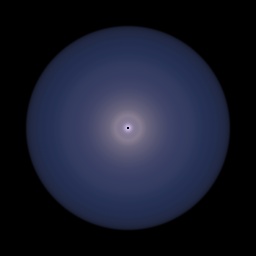}}\\
 & \multicolumn{2}{c|}{Observation Log-likelihood:} & \textcolor{red}{0.001} & 0.012 & 0.570 & 0.071 \\

\multirow{3}{*}{\rotatebox[origin=c]{90}{Violet Acrylic\hspace{0.7cm}}}& 
\raisebox{-0.5\height}{\includegraphics[width=\widthA]{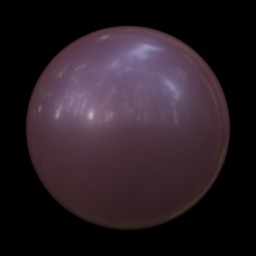}}&
\raisebox{-0.5\height}{\includegraphics[width=\widthA]{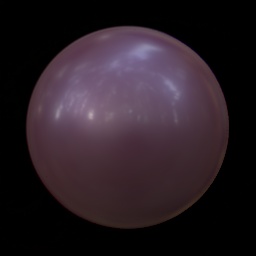}}&
\raisebox{-0.5\height}{\includegraphics[width=\widthA]{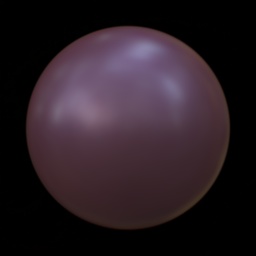}}&
\raisebox{-0.5\height}{\includegraphics[width=\widthA]{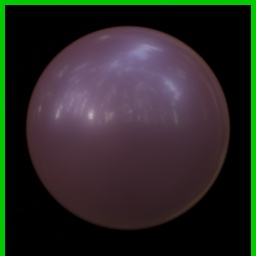}}&
\raisebox{-0.5\height}{\includegraphics[width=\widthA]{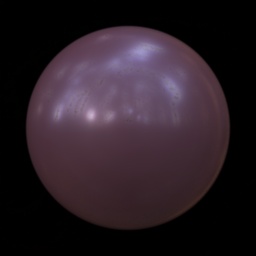}}&
\raisebox{-0.5\height}{\includegraphics[width=\widthA]{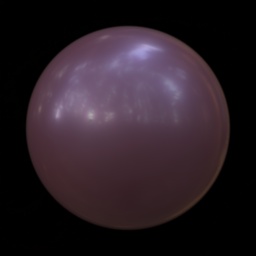}}\\
&
\raisebox{-0.5\height}{\includegraphics[width=\widthA]{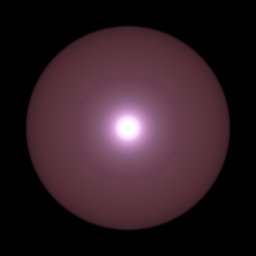}}&
\raisebox{-0.5\height}{\includegraphics[width=\widthA]{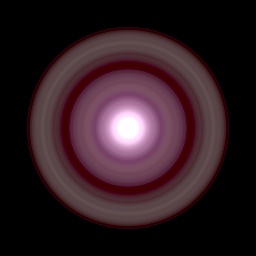}}&
\raisebox{-0.5\height}{\includegraphics[width=\widthA]{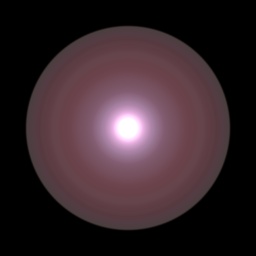}}&
\raisebox{-0.5\height}{\includegraphics[width=\widthA]{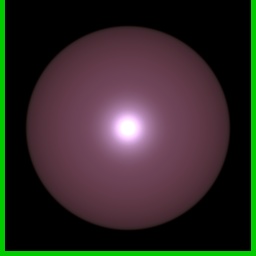}}&
\raisebox{-0.5\height}{\includegraphics[width=\widthA]{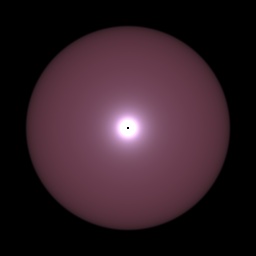}}&
\raisebox{-0.5\height}{\includegraphics[width=\widthA]{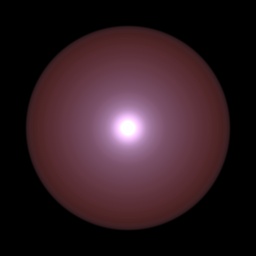}}\\
 & \multicolumn{2}{c|}{Observation Log-likelihood:} & 0.666 & \textcolor{red}{0.141} & 2.952 & 0.487 \\

\multirow{3}{*}{\rotatebox[origin=c]{90}{Chrome Steel\hspace{0.7cm}}}& 
\raisebox{-0.5\height}{\includegraphics[width=\widthA]{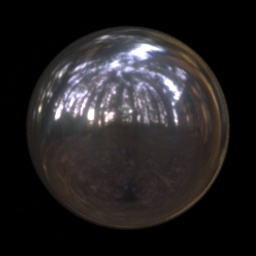}}&
\raisebox{-0.5\height}{\includegraphics[width=\widthA]{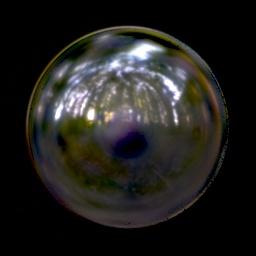}}&
\raisebox{-0.5\height}{\includegraphics[width=\widthA]{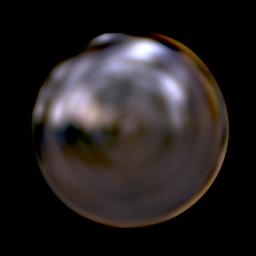}}&
\raisebox{-0.5\height}{\includegraphics[width=\widthA]{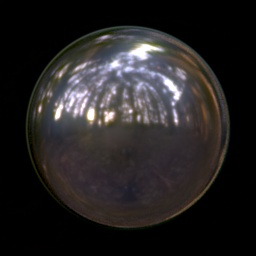}}&
\raisebox{-0.5\height}{\includegraphics[width=\widthA]{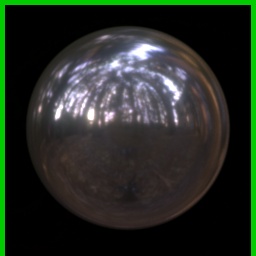}}&
\raisebox{-0.5\height}{\includegraphics[width=\widthA]{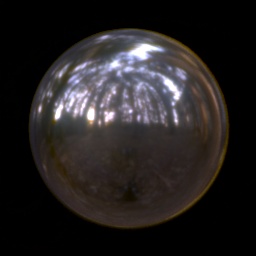}}\\
&
\raisebox{-0.5\height}{\includegraphics[width=\widthA]{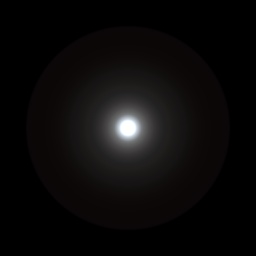}}&
\raisebox{-0.5\height}{\includegraphics[width=\widthA]{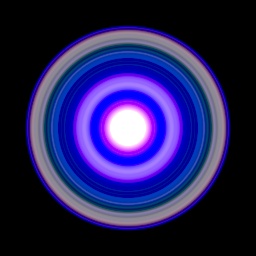}}&
\raisebox{-0.5\height}{\includegraphics[width=\widthA]{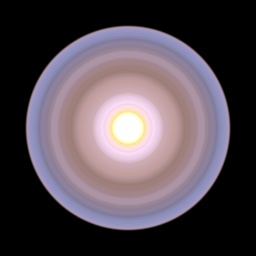}}&
\raisebox{-0.5\height}{\includegraphics[width=\widthA]{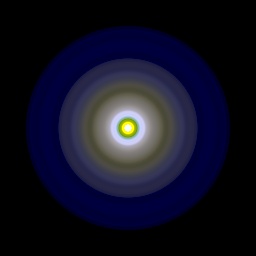}}&
\raisebox{-0.5\height}{\includegraphics[width=\widthA]{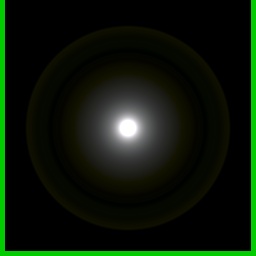}}&
\raisebox{-0.5\height}{\includegraphics[width=\widthA]{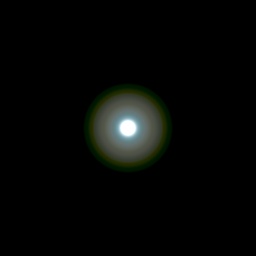}}\\
 & \multicolumn{2}{c|}{Observation Log-likelihood:} & 304.217 & 27.631 & \textcolor{red}{19.580} & 35.744 \\

\multirow{3}{*}{\rotatebox[origin=c]{90}{Red Metallic Paint\hspace{0.7cm}}}& 
\raisebox{-0.5\height}{\includegraphics[width=\widthA]{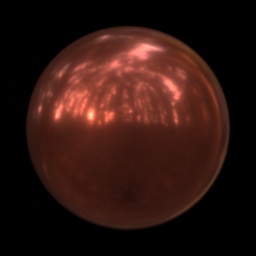}}&
\raisebox{-0.5\height}{\includegraphics[width=\widthA]{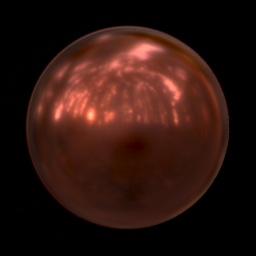}}&
\raisebox{-0.5\height}{\includegraphics[width=\widthA]{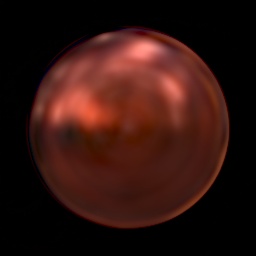}}&
\raisebox{-0.5\height}{\includegraphics[width=\widthA]{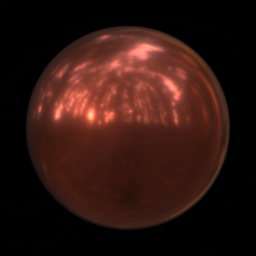}}&
\raisebox{-0.5\height}{\includegraphics[width=\widthA]{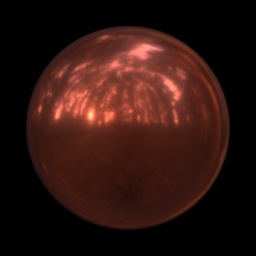}}&
\raisebox{-0.5\height}{\includegraphics[width=\widthA]{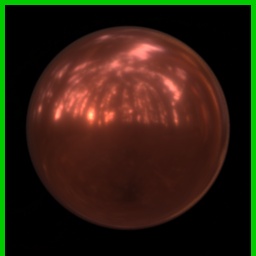}}\\
&
\raisebox{-0.5\height}{\includegraphics[width=\widthA]{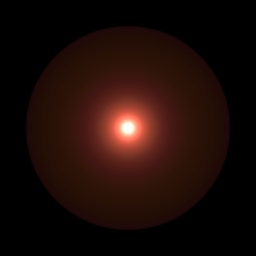}}&
\raisebox{-0.5\height}{\includegraphics[width=\widthA]{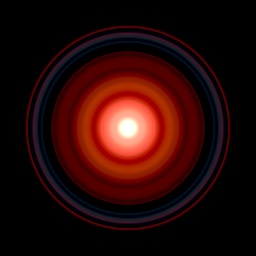}}&
\raisebox{-0.5\height}{\includegraphics[width=\widthA]{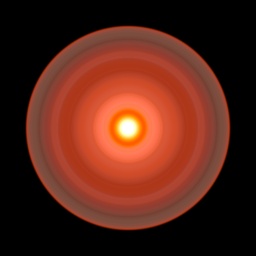}}&
\raisebox{-0.5\height}{\includegraphics[width=\widthA]{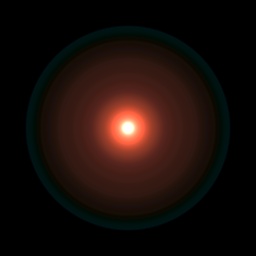}}&
\raisebox{-0.5\height}{\includegraphics[width=\widthA]{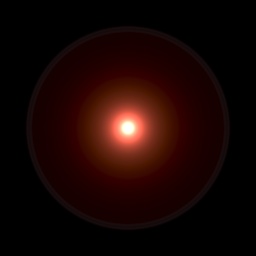}}&
\raisebox{-0.5\height}{\includegraphics[width=\widthA]{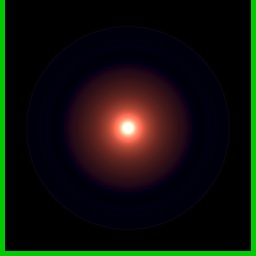}}\\
 & \multicolumn{2}{c|}{Observation Log-likelihood:} & 53.195 & 1.484 & 5.391 & \textcolor{red}{1.205} \\
\end{tabular}
}
\caption{Reconstructions for each material class for $4$ selected
  materials observed under the \emph{Uffizi Gallery} light probe, and
  revisualized under the \emph{Eucalyptus Grove} light probe and
  directional lighting. We list the log-likelihood error on the
  observations, and mark the best solution.  In addition we provide a
  comparison against a naive linear least squares reconstruction with
  the full MERL BRDFs.}
\label{fig:clusterimages}
\vspace{-1cm}
\end{center}
\end{figure*}

\paragraph{Experiment Setup}
We demonstrate our method on simulated reflectance maps in order to
fully control all parameters.  We generate the reflectance maps under
natural lighting, by rendering a sphere lit by a light
probe~\cite{Debevec:1998:Probes} using Mitsuba~\cite{Jakob:2010:MPB};
as noted in \autoref{sec:overview}, we will directly use this rendered
image as a representation of the reflectance map. All generated images
are radiometrically linear, and we only tone map them for display.
All results shown in this paper were tone mapped by a simple gamma
$2.2$ correction and a virtual exposure (i.e., scale factor) of $1.0$;
all pixel values above 1.0 and below 0.0 are clipped to the respective
clipping values. We use the BRDFs in the MERL
database~\cite{Matusik:2003:DRM} for generating reflectance maps.  For
each MERL BRDF, we compute a novel Gaussian mixture model on the $297$
remaining MERL BRDFs (i.e., we exclude the basis BRDF corresponding to
any of the three color channels of the BRDF), and only use these $297$
MERL BRDFs for reconstruction.  Consequently, any reconstruction of a
BRDF from the MERL BRDF database is computed using a different set of
basis BRDFs. As noted in the prior sections, we compute the Gaussian
mixture model on a $N=4$ dimensional reduced space, and use $K=4$
Gaussians in the mixture model.  All reconstructions are generated
with a fixed balancing factor $\lambda = 0.5$.

\paragraph{Reconstruction Results}
\autoref{fig:fullpage} shows reconstructions of $7$ selected materials
under two different light probes (i.e., \emph{Eucalyptus Grove} and
\emph{Galileo's Tomb}). For each reconstruction (and the reference),
we show a visualization of the reference/reconstructed BRDF under a
natural lighting condition (i.e., \emph{Uffizi Gallery}; different
than the lighting condition under which the BRDF was reconstructed)
and a directional light (i.e., a slice of the BRDF for a single
incident direction for all outgoing directions).  These results show
that our method is able to reconstruct plausible BRDFs for a wide
range of materials from a reflectance map under natural lighting. We
refer to the supplemental material for the reconstructions under
different natural lighting conditions for all MERL BRDFs.

\begin{figure*}[t!]
\begin{center}
{\setlength{\tabcolsep}{0mm}\small
\def\widthA{0.120\linewidth}
\begin{tabular}{ p{0.35cm} c c| c c| c c| c c }
 & \multicolumn{2}{c|}{} & \multicolumn{2}{c|}{Linear Least} & \multicolumn{2}{c|}{Single Material} & \multicolumn{2}{c}{Multiple Material}\\
 & \multicolumn{2}{c|}{Reference} & \multicolumn{2}{c|}{Squares} & \multicolumn{2}{c|}{Class Reconstruction} & \multicolumn{2}{c}{Class Reconstruction}\\

{\rotatebox[origin=c]{90}{Chrome Steel}}&
\raisebox{-0.5\height}{\includegraphics[width=\widthA]{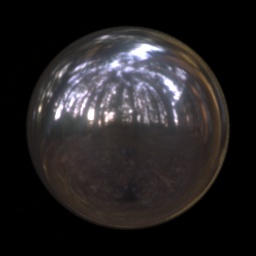}}&
\raisebox{-0.5\height}{\includegraphics[width=\widthA]{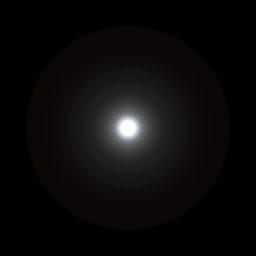}}&
\raisebox{-0.5\height}{\includegraphics[width=\widthA]{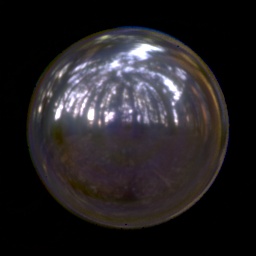}}&
\raisebox{-0.5\height}{\includegraphics[width=\widthA]{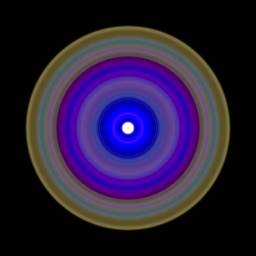}}&
\raisebox{-0.5\height}{\includegraphics[width=\widthA]{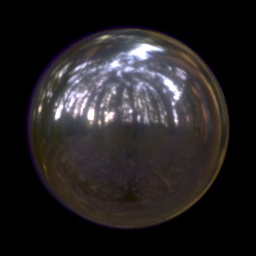}}&
\raisebox{-0.5\height}{\includegraphics[width=\widthA]{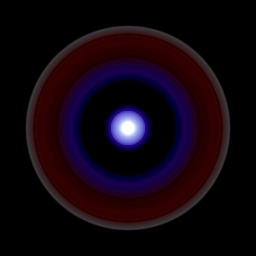}}&
\raisebox{-0.5\height}{\includegraphics[width=\widthA]{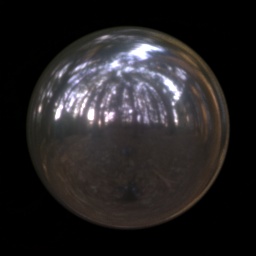}}&
\raisebox{-0.5\height}{\includegraphics[width=\widthA]{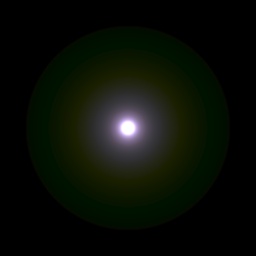}}\\

{\rotatebox[origin=c]{90}{Brown Fabric}}&
\raisebox{-0.5\height}{\includegraphics[width=\widthA]{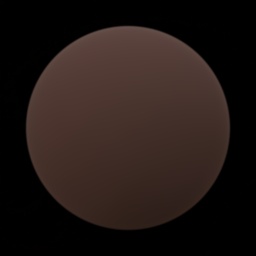}}&
\raisebox{-0.5\height}{\includegraphics[width=\widthA]{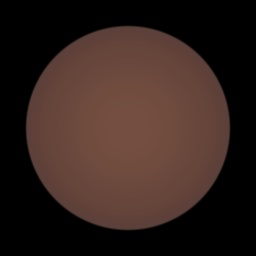}}&
\raisebox{-0.5\height}{\includegraphics[width=\widthA]{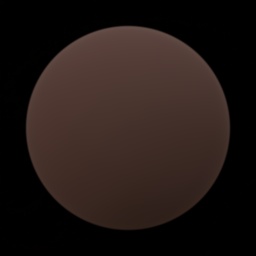}}&
\raisebox{-0.5\height}{\includegraphics[width=\widthA]{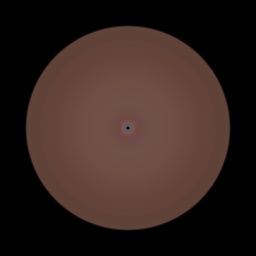}}&
\raisebox{-0.5\height}{\includegraphics[width=\widthA]{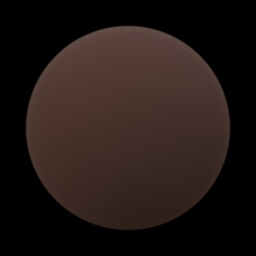}}&
\raisebox{-0.5\height}{\includegraphics[width=\widthA]{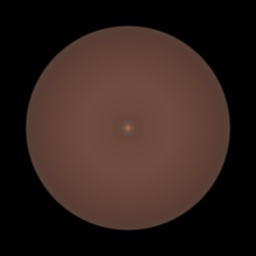}}&
\raisebox{-0.5\height}{\includegraphics[width=\widthA]{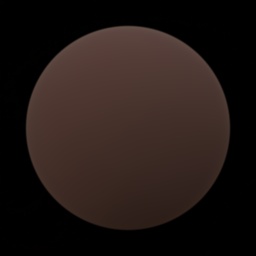}}&
\raisebox{-0.5\height}{\includegraphics[width=\widthA]{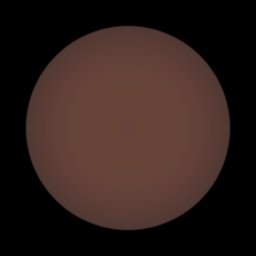}}\\

{\rotatebox[origin=c]{90}{Special Walnut}}&
\raisebox{-0.5\height}{\includegraphics[width=\widthA]{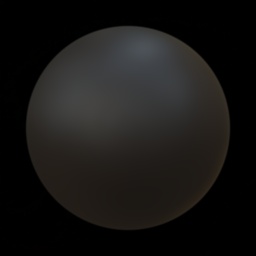}}&
\raisebox{-0.5\height}{\includegraphics[width=\widthA]{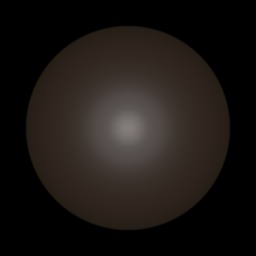}}&
\raisebox{-0.5\height}{\includegraphics[width=\widthA]{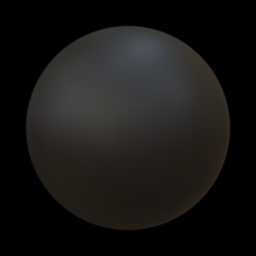}}&
\raisebox{-0.5\height}{\includegraphics[width=\widthA]{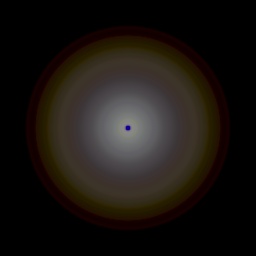}}&
\raisebox{-0.5\height}{\includegraphics[width=\widthA]{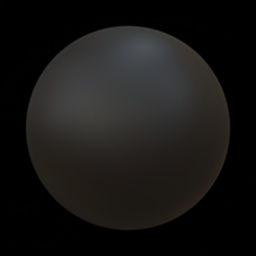}}&
\raisebox{-0.5\height}{\includegraphics[width=\widthA]{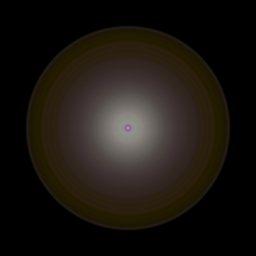}}&
\raisebox{-0.5\height}{\includegraphics[width=\widthA]{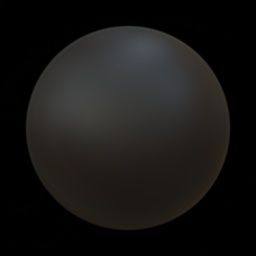}}&
\raisebox{-0.5\height}{\includegraphics[width=\widthA]{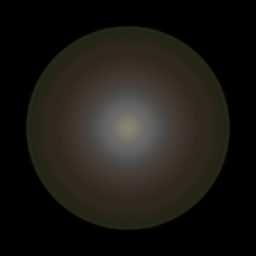}}\\

{\rotatebox[origin=c]{90}{Yellow Paint}}&

\raisebox{-0.5\height}{\includegraphics[width=\widthA]{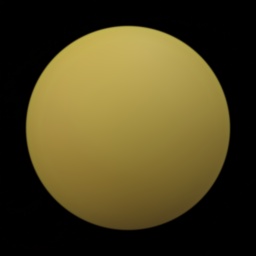}}&
\raisebox{-0.5\height}{\includegraphics[width=\widthA]{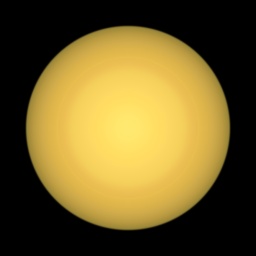}}&
\raisebox{-0.5\height}{\includegraphics[width=\widthA]{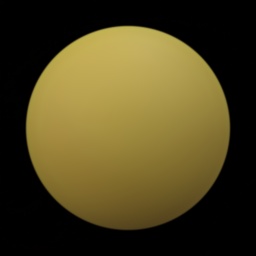}}&
\raisebox{-0.5\height}{\includegraphics[width=\widthA]{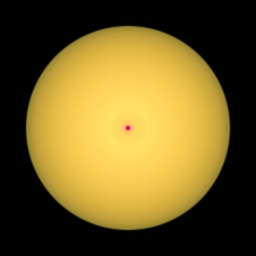}}&
\raisebox{-0.5\height}{\includegraphics[width=\widthA]{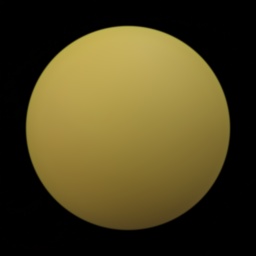}}&
\raisebox{-0.5\height}{\includegraphics[width=\widthA]{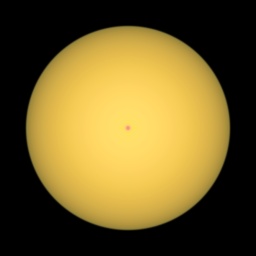}}&
\raisebox{-0.5\height}{\includegraphics[width=\widthA]{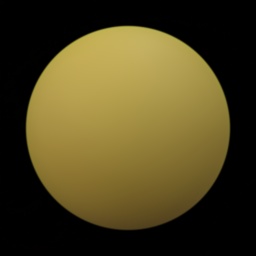}}&
\raisebox{-0.5\height}{\includegraphics[width=\widthA]{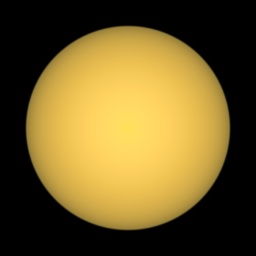}}\\
\end{tabular}
}
\caption{Comparison between naive linear least squares data-driven
  BRDF reconstruction, single material class reconstruction, and our
  multiple material class reconstruction.  For diffuse-like materials,
  both the linear least squares and single material class
  reconstructions exhibit a central ``spike'' visible under the
  directional lighting. For metals, strong ringing artifacts can be
  observed for both the linear least squares and single material class
  solutions.}
\label{fig:comparison}
\end{center}
\end{figure*}

\paragraph{Per-Material Class Reconstruction}
\autoref{fig:clusterimages} illustrates, for a selection of $4$
materials, reconstructed under the~\emph{Uf\-fi\-zi Gal\-le\-ry} light
probe, that the reconstructions per material class are different, and
that depending on the material a different class' reconstruction is
selected. We show a visualization of the reference BRDF and the
reconstructions per cluster under a natural lighting condition (i.e.,
\emph{Eucalyptus Grove}) and a directional light. We also list the
log-likelihood of the observation given the BRDF
(\autoref{eq:dataterm}) below each cluster, and mark the final
selected solution (i.e., minimum).  For reference, we also show the
linear least squares solution: $\argmin_\w ||\O\w - \o||^2$.  As
expected this yields the lowest reconstruction error (since it
explicitely optimizes for this). However, the linear least squares
solution does not always yield a plausible result when visualized
under a different lighting condition. This is not only clearly visible
under the directional light source, but also under other natural
lighting conditions other than the original observed lighting (e.g.,
the black spot in the center of the visualizations under the
\emph{Eucalyptus Grove} light probe for \emph{Steel} and \emph{Red
  Metallic Paint}).  Furthermore, we observe that not all clusters'
reconstructions appear to be plausible. However, the selection process
tends to pick the most plausible reconstruction.

\paragraph{Comparison: Single Material Class Reconstruction}
To gain insight in the importance of reconstructing the BRDF per
material class, we compare the reconstruction quality of the BRDF from
a single material class to our multi-material class solution
(\autoref{fig:comparison}).  Our results demonstrate that using a
single material class improves on a naive linear least
squares. However, our solution with multiple material classes
outperforms the single material class case.  Note that we optimized
$\lambda$ for the single cluster case to produce an as optimal result
given the lighting conditions. In this case we reconstructed the BRDF
under \emph{Grace Cathedral} lighting using a $\lambda = 0.5$ for the
single cluster case.  Note, that the single material class
reconstruction (\autoref{eq:lls}) is similar to
Nielsen~\etal's~\cite{Nielsen:2015:OMB} method, without applying a
non-linear encoding of the BRDF. Additional minor differences are that
Nielsen~\etal subtract the median instead of the mean before
computing the linear least squares and assume a unit standard
deviation.  Furthermore, the single class reconstruction is also
similar to Romeiro~\etal's~\cite{Romeiro:2008:PR} method, using a
linear data-driven BRDF model instead of the bivariate model. Since we
a-priori assume a linear BRDF model, we want to explore the
differences between the reconstruction methods, not the BRDF model
representations.

In general, we found that overall our method outperforms a single
material class reconstruction.  The single material reconstruction
tends to work equally well on phenolic and plastic materials as these
are similar in BRDF shape to the mean material.  However, the single
material class reconstruction fails for diffuse and metal-like
materials.  While less strong than for the naive least squares, for
diffuse materials we can observe a central ``spike'' visible under the
directional lighting.  For the metal-like materials, we typically
observe strong ringing artifacts.

\begin{figure*}[t!]
\begin{center}
{\setlength{\tabcolsep}{0mm}\small
\def\widthA{0.125\linewidth}

\begin{tabular}{ p{0.35cm} c| c| c c c c }
 &           & Linear Least & \multicolumn{4}{c}{Material Class}\\
 & Reference & Squares & Diffuse & Plastics/Phenolics & Metals & Spec. Plastic/Paints \\

\multirow{4}{*}{\rotatebox[origin=c]{90}{Orange Foam Ball\hspace{0.7cm}}}& 
\raisebox{-0.5\height}{\includegraphics[width=\widthA]{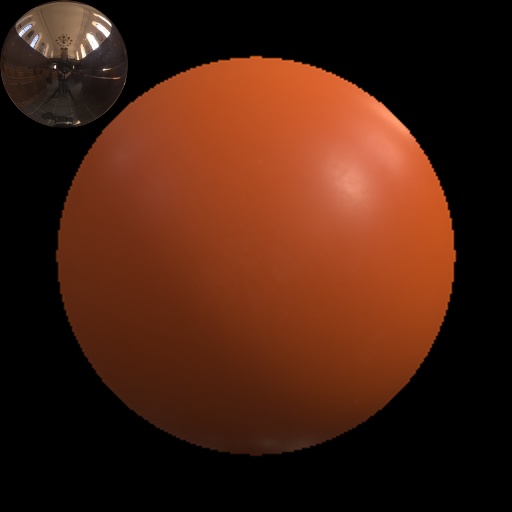}}&
\raisebox{-0.5\height}{\includegraphics[width=\widthA]{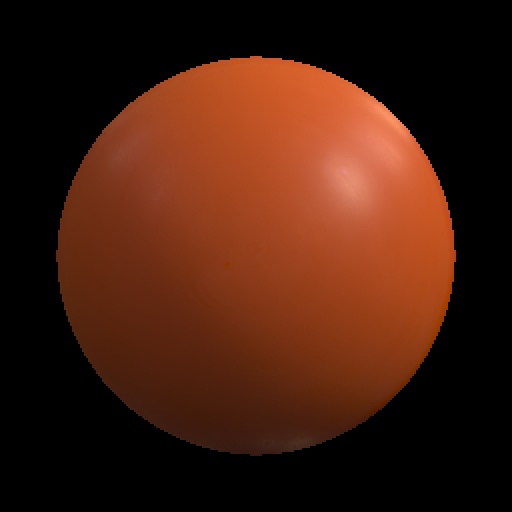}}&
\raisebox{-0.5\height}{\includegraphics[width=\widthA]{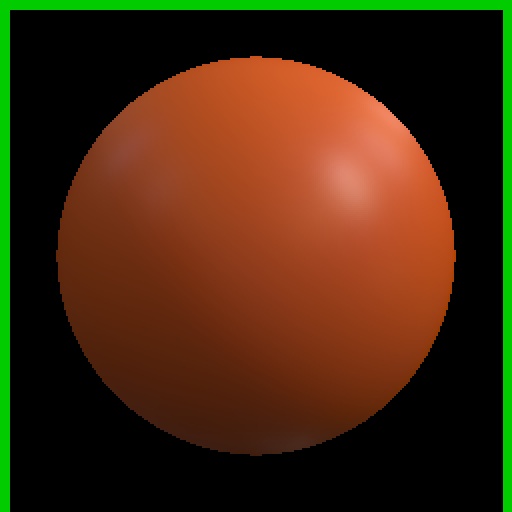}}&
\raisebox{-0.5\height}{\includegraphics[width=\widthA]{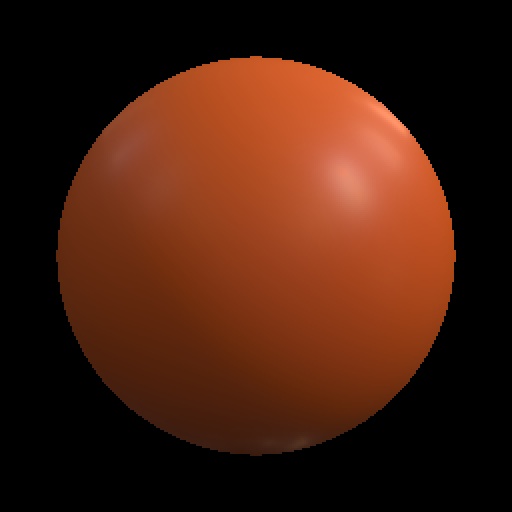}}&
\raisebox{-0.5\height}{\includegraphics[width=\widthA]{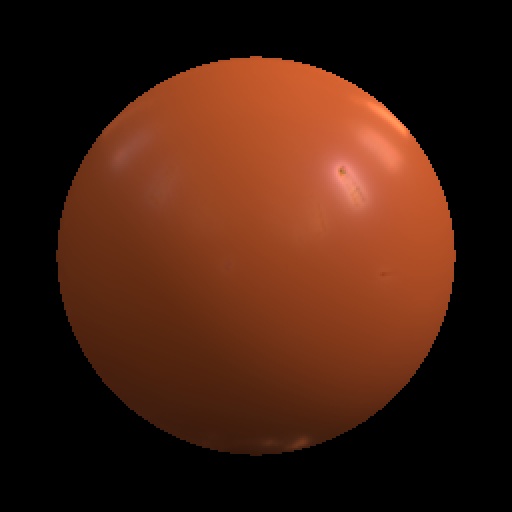}}&
\raisebox{-0.5\height}{\includegraphics[width=\widthA]{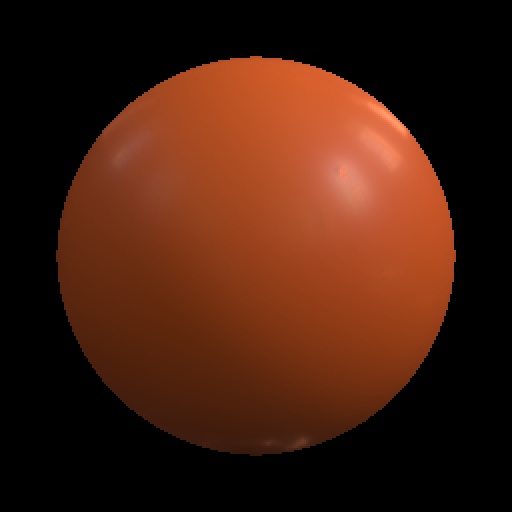}}\\
&
\raisebox{-0.5\height}{\includegraphics[width=\widthA]{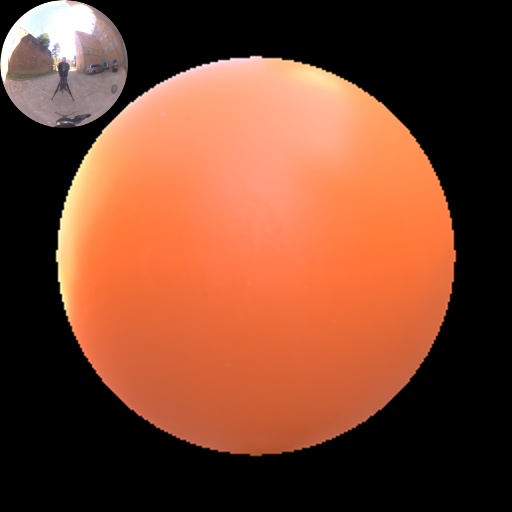}}&
\raisebox{-0.5\height}{\includegraphics[width=\widthA]{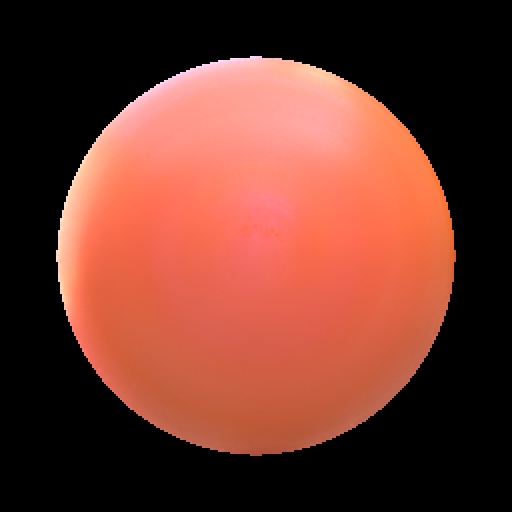}}&
\raisebox{-0.5\height}{\includegraphics[width=\widthA]{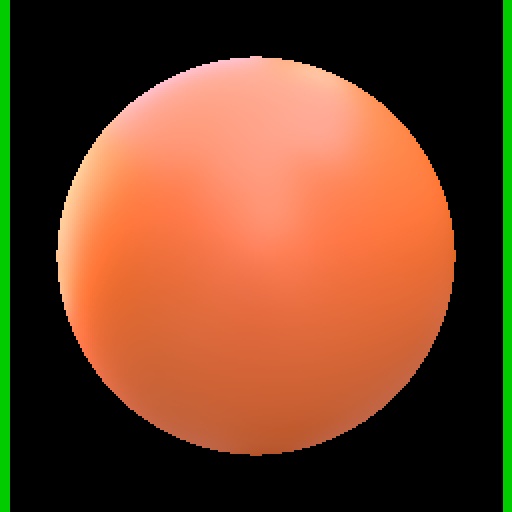}}&
\raisebox{-0.5\height}{\includegraphics[width=\widthA]{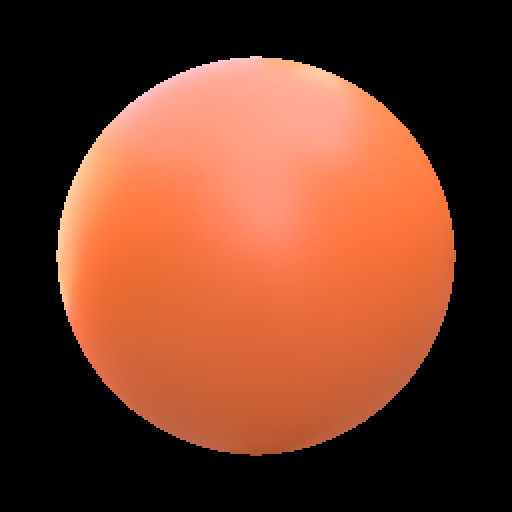}}&
\raisebox{-0.5\height}{\includegraphics[width=\widthA]{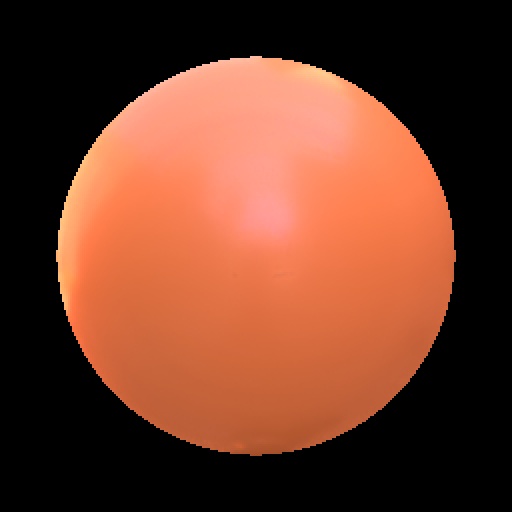}}&
\raisebox{-0.5\height}{\includegraphics[width=\widthA]{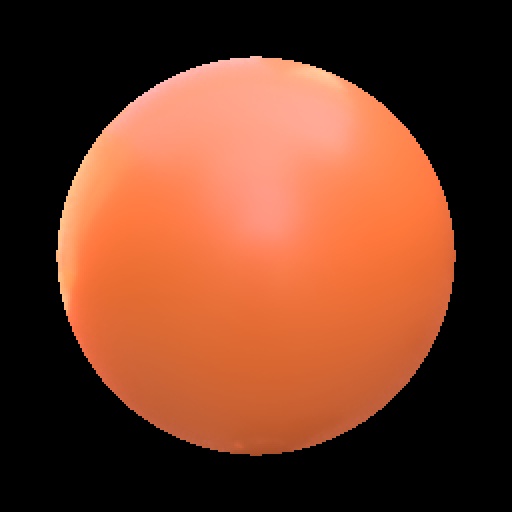}}\\
&
\raisebox{-0.5\height}{\includegraphics[width=\widthA]{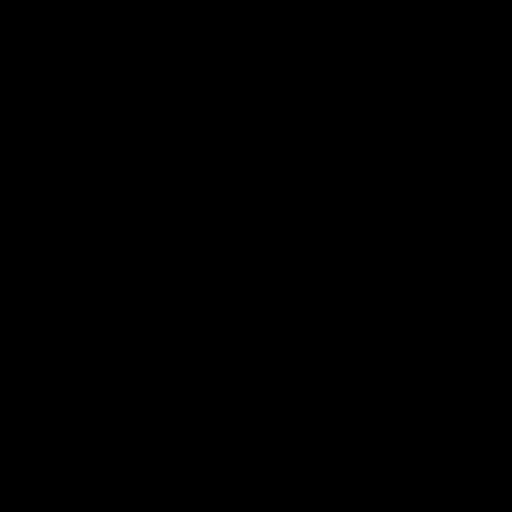}}&
\raisebox{-0.5\height}{\includegraphics[width=\widthA]{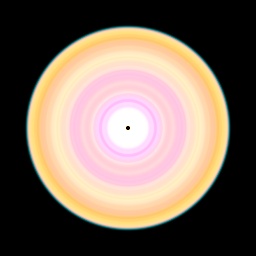}}&
\raisebox{-0.5\height}{\includegraphics[width=\widthA]{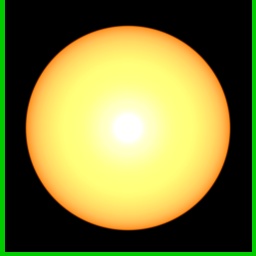}}&
\raisebox{-0.5\height}{\includegraphics[width=\widthA]{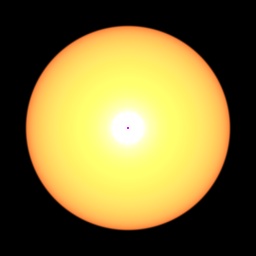}}&
\raisebox{-0.5\height}{\includegraphics[width=\widthA]{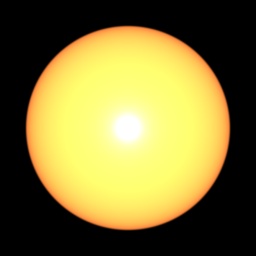}}&
\raisebox{-0.5\height}{\includegraphics[width=\widthA]{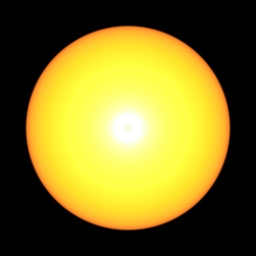}}\\

 & \multicolumn{2}{c|}{Observation Log-likelihood:} & \textcolor{red}{0.5963} & 0.6614 & 7.7250 & 1.5751 \\

\multirow{4}{*}{\rotatebox[origin=c]{90}{Blue Plastic\hspace{0.7cm}}}& 
\raisebox{-0.5\height}{\includegraphics[width=\widthA]{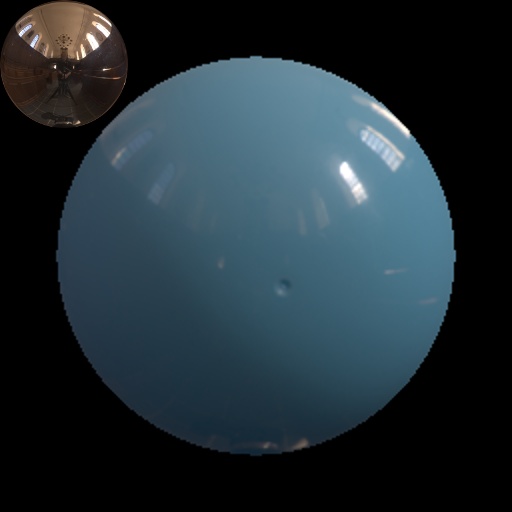}}&
\raisebox{-0.5\height}{\includegraphics[width=\widthA]{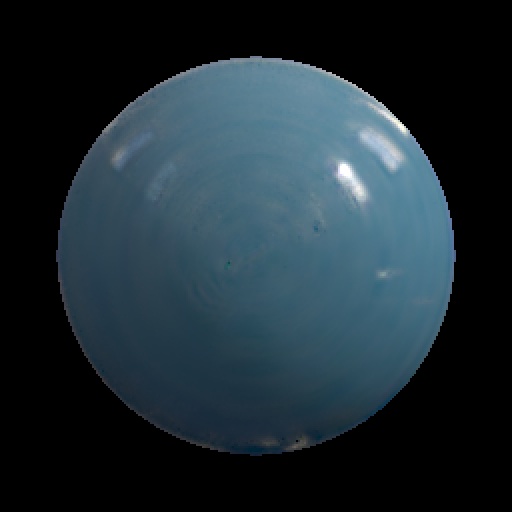}}&
\raisebox{-0.5\height}{\includegraphics[width=\widthA]{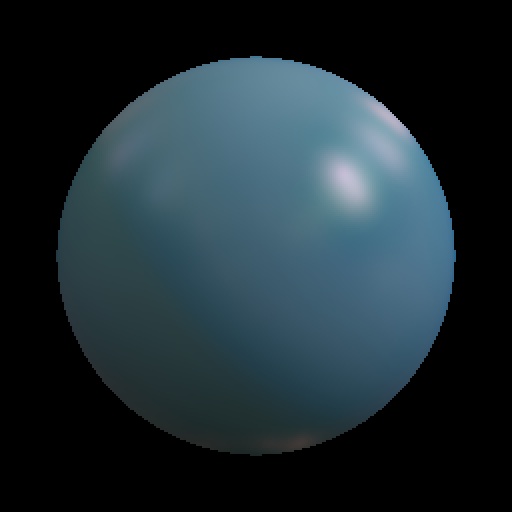}}&
\raisebox{-0.5\height}{\includegraphics[width=\widthA]{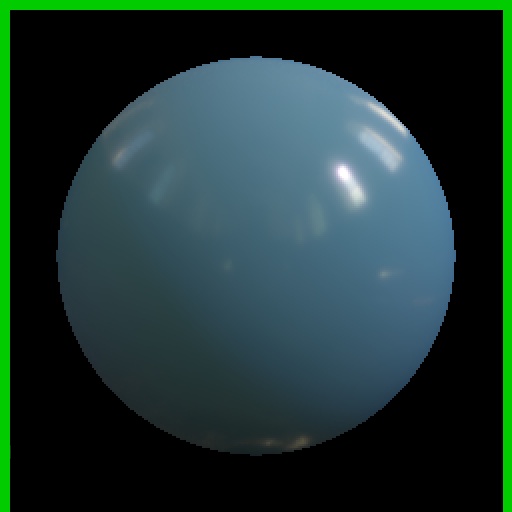}}&
\raisebox{-0.5\height}{\includegraphics[width=\widthA]{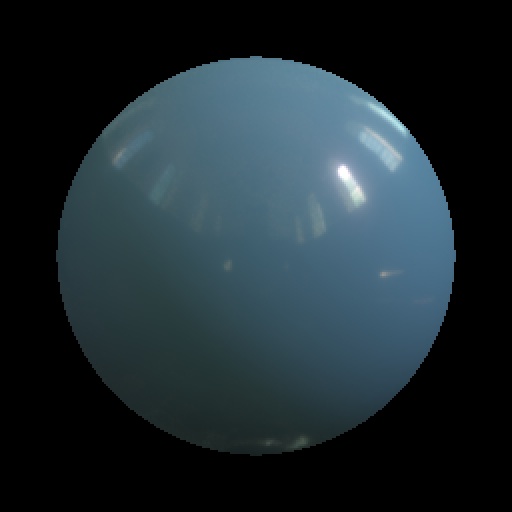}}&
\raisebox{-0.5\height}{\includegraphics[width=\widthA]{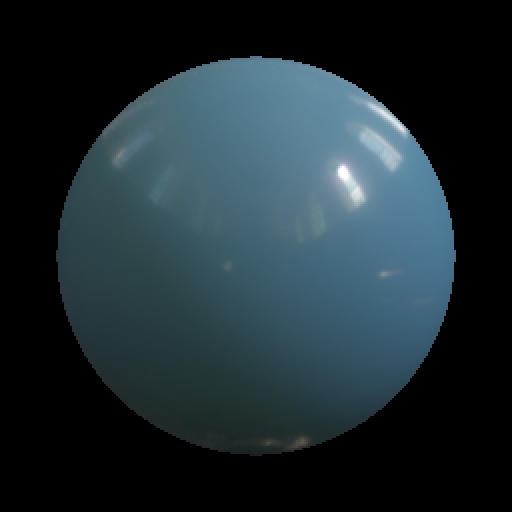}}\\
&
\raisebox{-0.5\height}{\includegraphics[width=\widthA]{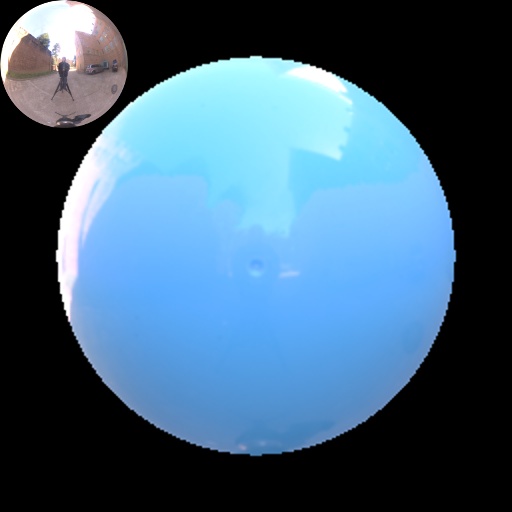}}&
\raisebox{-0.5\height}{\includegraphics[width=\widthA]{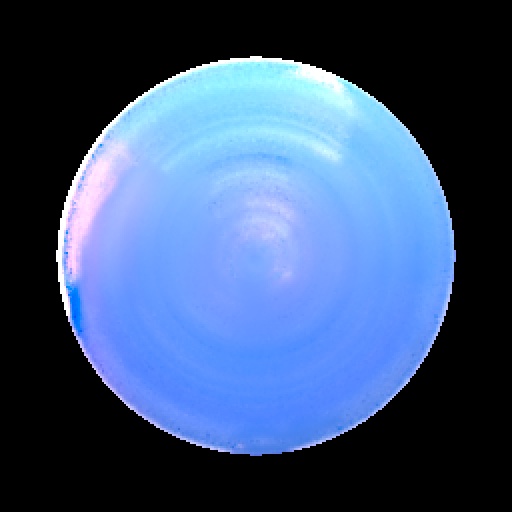}}&
\raisebox{-0.5\height}{\includegraphics[width=\widthA]{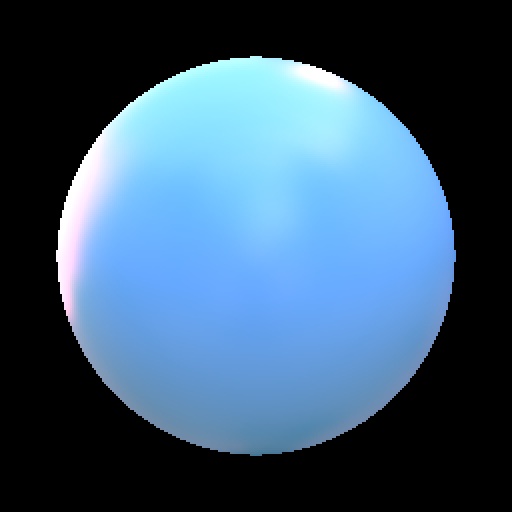}}&
\raisebox{-0.5\height}{\includegraphics[width=\widthA]{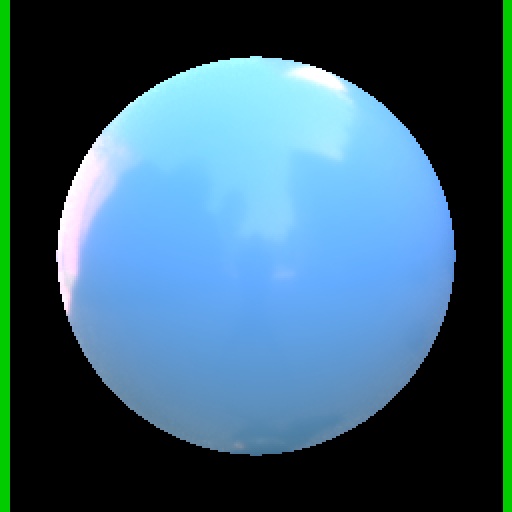}}&
\raisebox{-0.5\height}{\includegraphics[width=\widthA]{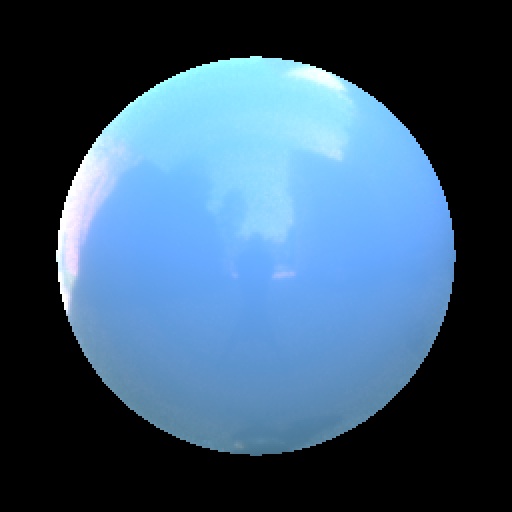}}&
\raisebox{-0.5\height}{\includegraphics[width=\widthA]{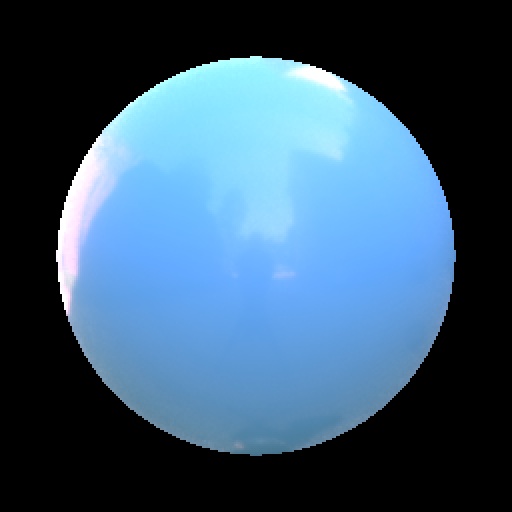}}\\
&
\raisebox{-0.5\height}{\includegraphics[width=\widthA]{figures/fig_realdata/black.jpg}}&
\raisebox{-0.5\height}{\includegraphics[width=\widthA]{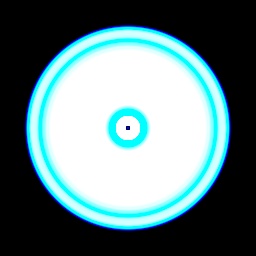}}&
\raisebox{-0.5\height}{\includegraphics[width=\widthA]{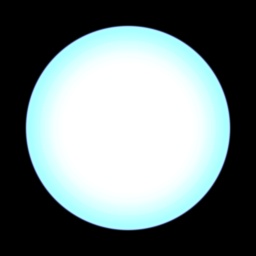}}&
\raisebox{-0.5\height}{\includegraphics[width=\widthA]{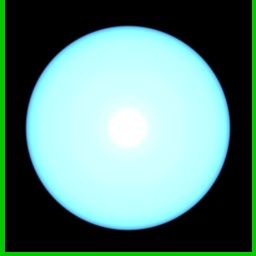}}&
\raisebox{-0.5\height}{\includegraphics[width=\widthA]{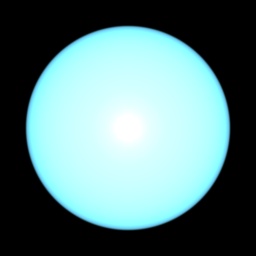}}&
\raisebox{-0.5\height}{\includegraphics[width=\widthA]{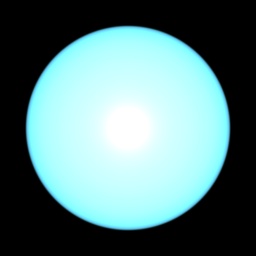}}\\

 & \multicolumn{2}{c|}{Observation Log-likelihood:} & 24.8899 & \textcolor{red}{11.8852} & 12.8056 & 12.8498 \\

\multirow{4}{*}{\rotatebox[origin=c]{90}{Dark Bronze\hspace{0.7cm}}}& 
\raisebox{-0.5\height}{\includegraphics[width=\widthA]{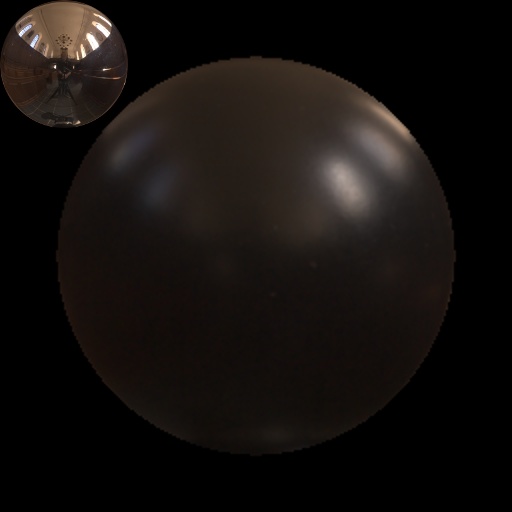}}&
\raisebox{-0.5\height}{\includegraphics[width=\widthA]{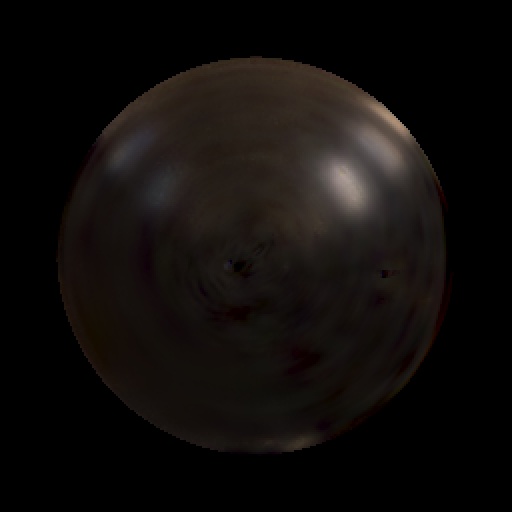}}&
\raisebox{-0.5\height}{\includegraphics[width=\widthA]{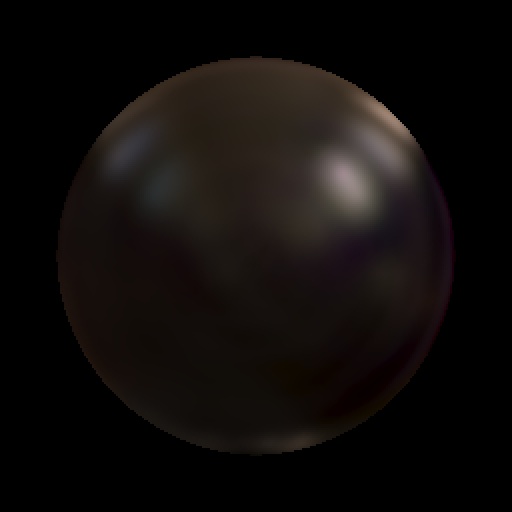}}&
\raisebox{-0.5\height}{\includegraphics[width=\widthA]{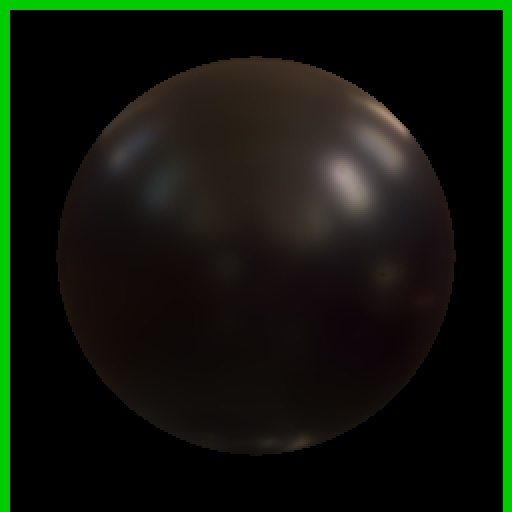}}&
\raisebox{-0.5\height}{\includegraphics[width=\widthA]{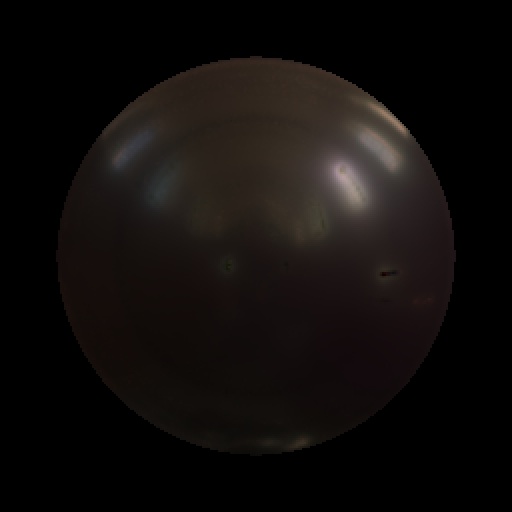}}&
\raisebox{-0.5\height}{\includegraphics[width=\widthA]{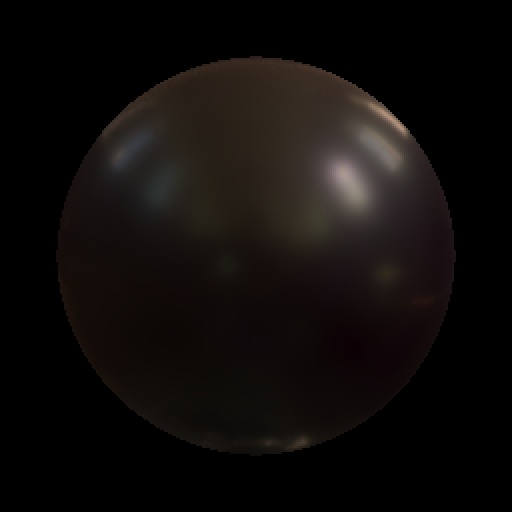}}\\
&
\raisebox{-0.5\height}{\includegraphics[width=\widthA]{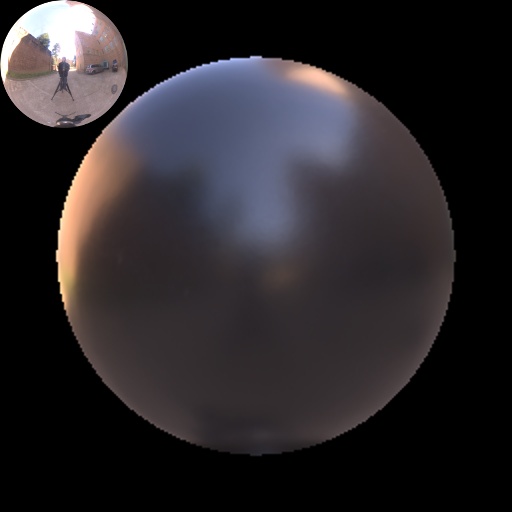}}&
\raisebox{-0.5\height}{\includegraphics[width=\widthA]{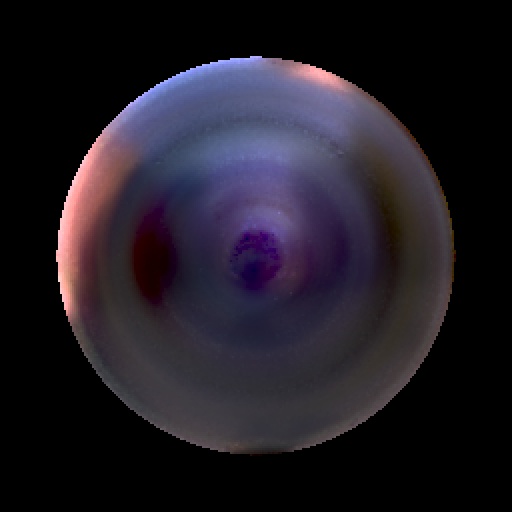}}&
\raisebox{-0.5\height}{\includegraphics[width=\widthA]{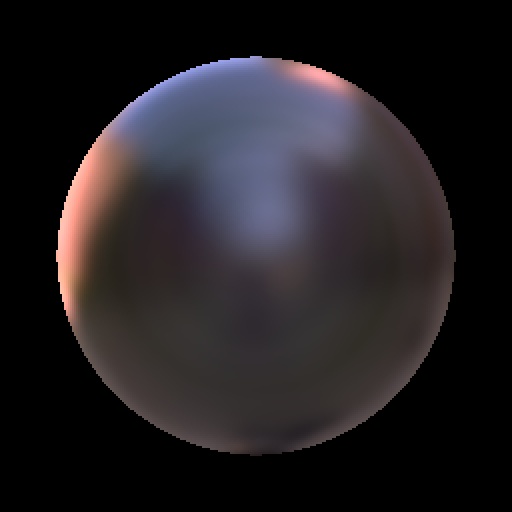}}&
\raisebox{-0.5\height}{\includegraphics[width=\widthA]{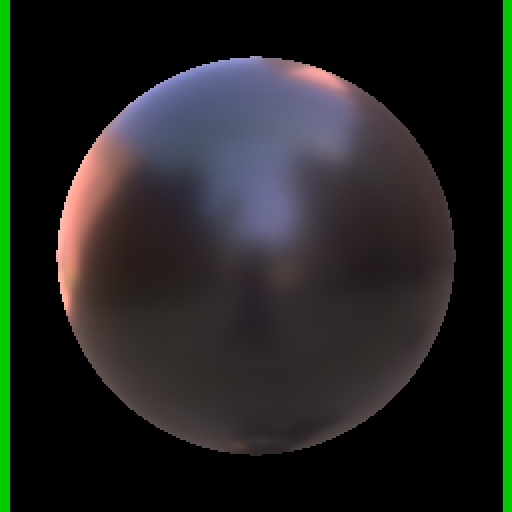}}&
\raisebox{-0.5\height}{\includegraphics[width=\widthA]{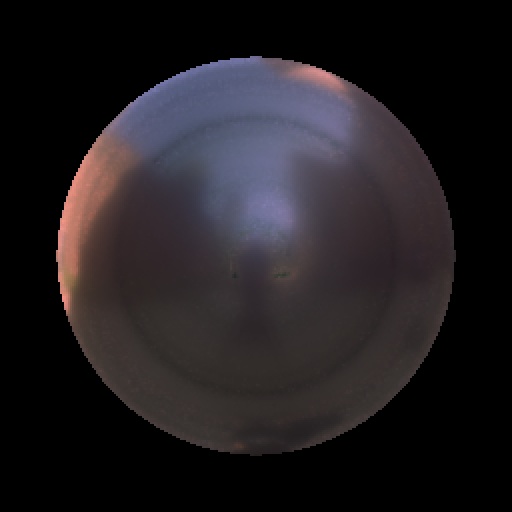}}&
\raisebox{-0.5\height}{\includegraphics[width=\widthA]{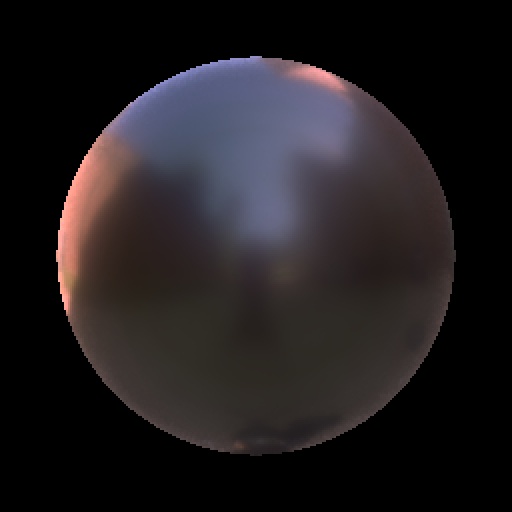}}\\
&
\raisebox{-0.5\height}{\includegraphics[width=\widthA]{figures/fig_realdata/black.jpg}}&
\raisebox{-0.5\height}{\includegraphics[width=\widthA]{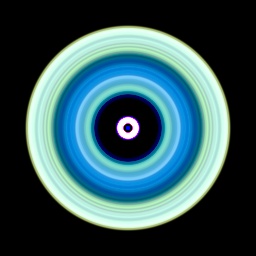}}&
\raisebox{-0.5\height}{\includegraphics[width=\widthA]{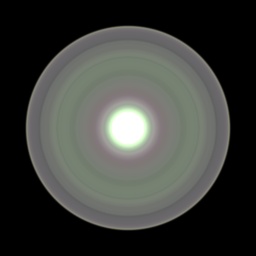}}&
\raisebox{-0.5\height}{\includegraphics[width=\widthA]{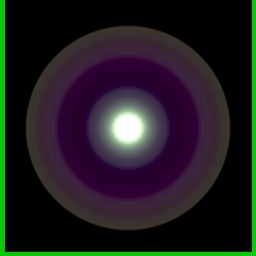}}&
\raisebox{-0.5\height}{\includegraphics[width=\widthA]{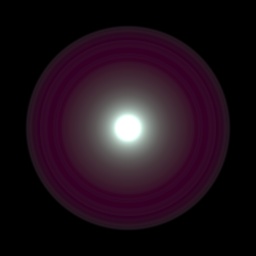}}&
\raisebox{-0.5\height}{\includegraphics[width=\widthA]{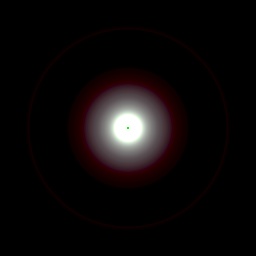}}\\

 & \multicolumn{2}{c|}{Observation Log-likelihood:} & 0.7201 & \textcolor{red}{0.4698} & 1.1173 & 0.6966 \\

\end{tabular}
}

\caption{Reconstructions for each material class for $3$ captured
  materials observed under indoor natural lighting, and revisualized
  under outdoor natural lighting and directional lighting. For each
  natural lighting condition we also provide a reference
  photograph. We list the log-likelihood error, and mark the best
  solution.  In addition we provide a comparison against a naive
  linear least squares reconstruction.}
\label{fig:realdata}
\vspace{-1cm}
\end{center}
\end{figure*}

\paragraph{Captured Reflectance Map Validation}
To validate our method on other materials than the MERL BRDF database,
we performed the following proof-of-concept experiment.  We acquired
three spheres with different materials (i.e., \emph{Dense Orange
Foam}, \emph{Blue Plastic}, and \emph{Dark Bronze}) under two
different natural lighting environments shown in the insets. Next, we
estimate data-driven BRDF parameters under the indoor \emph{Chapel}
lighting, and rerender the sphere under the outdoor lighting.  We mask
out any measured reflectance values that deviate from the expected
measurement conditions (e.g., the dimple on the \emph{Blue Plastic},
and the near field reflection from the stand). As can be seen
in~\autoref{fig:realdata}, the rerendered reflectance maps closely
resemble the acquired reference maps.  Note, the \emph{Dark Bronze}
material exhibits anisotropic reflectance which adversely impacts the
reconstruction. Nevertheless, the reconstructed BRDF remains
plausible. For reference, we also include a least squares data-driven
BRDF reconstruction. In addition, we also show visualizations of the
reconstructed BRDFs lit by a directional light source to better
demonstrate the plausibility of the reconstructions.

\paragraph{Discussion}
While our selection criterion does in the majority of cases select the
best reconstruction from the different material classes, we found that
in a few cases cases it does not select the best reconstruction, and a
better reconstruction can be observed in another material class.
Ideally, the selection criterion should not only include the data
term~\autoref{eq:dataterm}, but evaluate the full non-linear MAP
estimation loss (\autoref{eq:nonlinloss}).  We observe that for cases
where our current selection criterion prefers a suboptimal solution,
that the accompanying likelihood term $\P(\hat{\U}^T \fr)$ is
relatively large. However, a challenge is that the range of the
data-term and the likelihood cover a different range due to: (a) the
ommission of a standard deviation scale in the data-term, and (b) the
dimension reduction in the likelihood term.  Finding a good balancing
term is non-trivial and an interesting avenue for future research.

We currently used a $\lambda$ balancing factor of $0.5$ for all our
reconstructions.  This $\lambda$ is a compromise to produce the best
result over all materials.  Despite the material class and scene
dependent scale factor (\autoref{eq:lambda}), we observe that this
lambda terms tends to affect the \emph{``diffuse''} and
\emph{``plastics / phe\-nolics''} stronger, and the \emph{''metals''} and
\emph{``specular plastics / paints''} less.  These latter two material
classes exhibit not only a lower number of materials (for which we
compensate), but we can also observe in~\autoref{fig:plot} that they
are also spread out further. Consequently, the density of these
material classes is significantly lower.  This lower density implies
that the material class is very diverse in BRDF types and that the
MERL BRDF database does not densely sample these material
types. Taking in account this density difference is another
interesting avenue for future work.  In general, we find that
reconstructions from these material classes are less often selected.

\paragraph{Relation to Prior Work}
Matusik~\etal~\cite{Matusik:2003:DRM} showed that the log-encoded
BRDF space can be accurately modeled by a $45$D linear subspace and a
$15$D non-linear manifold.  While a linear model is computationally
more convenient, a non-linear model offers a tighter fit to the space
of BRDFs, and consequently, it contains less implausible BRDFs.  Our
Gaussian mixture based model can be seen as a piecewise linear
approximation of the non-linear manifold of BRDFs.  In contrast to
Matusik~\etal, we work directly on the space spanned by the basis
BRDF (i.e., without log-encoding).  However, as shown
in~\autoref{fig:plot} this manifold is highly non-linear too.  While
less tight than a full $15$D non-linear model, our Gaussian mixture
models strikes a balance between tightness and the ability to robustly
identify the piecewise linear subspace to which the observations under
natural lighting belong.

An implausible BRDF lies inside the linear subspace spanned by the
linear model, but outside the non-linear BRDF manifold.  Ideally, we
would like to bias these implausible solution towards the non-linear
manifold to obtain a more plausible solution.  Tikhonov regularization
biases the reconstruction towards a mean BRDF, assuming that the
solution is more plausible when closer to the mean.  However, such a
regularization is only efficient if the modeled space resembles a
hypersphere. Nielsen~\etal~\cite{Nielsen:2015:OMB} model the BRDF
space as a hypersphere by scaling the PCA basis BRDFs by the singular
values.  However, as noted before, the BRDF space is highly non-linear
and such a hypersphere is not a tight model.  Intuitively, biasing
the reconstruction of a diffuse material towards the mean or median
BRDF is suboptimal; the mean or median BRDF contains a rough specular
lobe.  Consequently, it is possible that biasing pushes the solution
towards a point away from the non-linear manifold. Our solution
represents the space of BRDFs as a sum of (rescaled) hyperspheres: the
per-material class linear likelihood term (\autoref{eq:biasterm})
biases the solution to the mean ($\mu'_j$) of the local hypersphere
(rescaled by $\Sigma'_j$).  Since each Gaussian subspace is more
tight, biasing towards the mean has a lower likelihood of ending away
from the non-linear manifold.

\section{Conclusion}
\label{sec:conclusion}

In this paper we presented a novel method for estimating the
parameters of a fully linear data-driven BRDF model from a reflectance
map under uncontrolled, but known, natural lighting. Our estimation
method does not require any non-linear optimization, and only requires
solving $4$ linear least squares problems. Our method requires modest
precomputations: a Gaussian mixture model clustering for the basis
BRDFs, and for each natural lighting conditions, renderings of each
basis material.  We demonstrated the accuracy and robustness of our
method on the MERL BRDF database, and validated our method on
real-world measurements.

For future work we would like to explore better selection criteria and
a per-material class $\lambda_j$ density correction factor.

\paragraph{Acknowledgments} This work was supported in part by NSF
grant IIS-1350323 and gifts from Google, Activision, and Nvidia.

{\small
\bibliographystyle{ieee} 
\bibliography{references}       

\begin{thebibliography}{10}\itemsep=-1pt

\bibitem{Bagher:2016:ANP}
M.~M. Bagher, J.~Snyder, and D.~Nowrouzezahrai.
\newblock A non-parametric factor microfacet model for isotropic brdfs.
\newblock {\em {ACM} Trans. Graph.}, 35(5):159:1--159:16, 2016.

\bibitem{Barron:2015:SIR}
J.~T. Barron and J.~Malik.
\newblock Shape, illumination, and reflectance from shading.
\newblock {\em IEEE PAMI}, 2015.

\bibitem{Debevec:1998:Probes}
P.~Debevec.
\newblock Light probe gallery.
\newblock \url{http://www.pauldebevec.com/Probes/}, 1998.

\bibitem{Dong:2014:ARS}
Y.~Dong, G.~Chen, P.~Peers, J.~Zhang, and X.~Tong.
\newblock Appearance-from-motion: Recovering spatially varying surface
  reflectance under unknown lighting.
\newblock {\em ACM Trans. Graph.}, 33(6):193:1--193:12, 2014.

\bibitem{Dorsey:2007:DMM}
J.~Dorsey, H.~Rushmeier, and F.~Sillion.
\newblock {\em Digital Modeling of Material Appearance}.
\newblock Morgan Kaufmann Publishers Inc., 2008.

\bibitem{Jakob:2010:MPB}
W.~Jakob.
\newblock Mitsuba: Physically based renderer.
\newblock \url{https://www.mitsuba-renderer.org}, 2010.

\bibitem{Li:2017:MSA}
X.~Li, Y.~Dong, P.~Peers, and X.~Tong.
\newblock Modeling surface appearance from a single photograph using
  self-augmented convolutional neural networks.
\newblock {\em ACM Trans. Graph.}, 36(4):45:1--45:11, July 2017.

\bibitem{Li:2018:MMS}
Z.~Li, K.~Sunkavalli, and M.~K. Chandraker.
\newblock Materials for masses: Svbrdf acquisition with a single mobile phone
  image.
\newblock In {\em ECCV}, 2018.

\bibitem{Li:2018:LRS}
Z.~Li, Z.~Xu, R.~Ramamoorthi, K.~Sunkavalli, and M.~Chandraker.
\newblock Learning to reconstruct shape and spatially-varying reflectance from
  a single image.
\newblock {\em ACM Trans. Graph.}, 37(6), Dec. 2018.

\bibitem{Lombardi:2016:RIR}
S.~Lombardi and K.~Nishino.
\newblock Reflectance and illumination recovery in the wild.
\newblock {\em IEEE PAMI}, 38(1):129--141, 2016.

\bibitem{Matusik:2003:DRM}
W.~Matusik, H.~Pfister, M.~Brand, and L.~McMillan.
\newblock A data-driven reflectance model.
\newblock {\em ACM Trans. Graph.}, 22(3):759--769, July 2003.

\bibitem{Matusik:2003:EIB}
W.~Matusik, H.~Pfister, M.~Brand, and L.~McMillan.
\newblock Efficient isotropic {BRDF} measurement.
\newblock In {\em Rendering Techniques}, pages 241--248, 2003.

\bibitem{Nielsen:2015:OMB}
J.~B. Nielsen, H.~W. Jensen, and R.~Ramamoorthi.
\newblock On optimal, minimal brdf sampling for reflectance acquisition.
\newblock {\em ACM Trans. Graph.}, 34(6), Oct. 2015.

\bibitem{Nishino:2011:DSB}
K.~Nishino and S.~Lombardi.
\newblock Directional statistics-based reflectance model for isotropic
  bidirectional reflectance distribution functions.
\newblock {\em J. Opt. Soc. Am. A}, 28(1):8--18, Jan 2011.

\bibitem{Oxholm:2016:SRE}
G.~Oxholm and K.~Nishino.
\newblock Shape and reflectance estimation in the wild.
\newblock {\em IEEE PAMI}, 38(2):376--389, Feb. 2016.

\bibitem{Palma:2012:SMS}
G.~Palma, M.~Callieri, M.~Dellepiane, and R.~Scopigno.
\newblock A statistical method for svbrdf approximation from video sequences in
  general lighting conditions.
\newblock {\em Comput. Graph. Forum}, 31(4):1491--1500, 2012.

\bibitem{Ramamoorthi:2001:SFI}
R.~Ramamoorthi and P.~Hanrahan.
\newblock A signal-processing framework for inverse rendering.
\newblock In {\em Proceedings of the 28th Annual Conference on Computer
  Graphics and Interactive Techniques}, SIGGRAPH '01, pages 117--128, 2001.

\bibitem{Rematas:2016:DRM}
K.~Rematas, T.~Ritschel, M.~Fritz, E.~Gavves, and T.~Tuytelaars.
\newblock Deep reflectance maps.
\newblock In {\em CVPR}, 2016.

\bibitem{Romeiro:2008:PR}
F.~Romeiro, Y.~Vasilyev, and T.~Zickler.
\newblock Passive reflectometry.
\newblock In {\em ECCV}, pages 859--872, 2008.

\bibitem{Romeiro:2010:BR}
F.~Romeiro and T.~Zickler.
\newblock Blind reflectometry.
\newblock In {\em ECCV}, pages 45--58, 2010.

\bibitem{Walter:2007:MMR}
B.~Walter, S.~R. Marschner, H.~Li, and K.~E. Torrance.
\newblock Microfacet models for refraction through rough surfaces.
\newblock In {\em Rendering Techniques}, pages 195--206, 2007.

\bibitem{Ward:1992:MMA}
G.~J. Ward.
\newblock Measuring and modeling anisotropic reflection.
\newblock {\em SIGGRAPH Comput. Graph.}, 26(2):265--272, 1992.

\bibitem{Weinmann:2015:AGR}
M.~Weinmann and R.~Klein.
\newblock Advances in geometry and reflectance acquisition.
\newblock In {\em ACM SIGGRAPH Asia, Course Notes}, 2015.

\bibitem{Xia:2016:RSS}
R.~Xia, Y.~Dong, P.~Peers, and X.~Tong.
\newblock Recovering shape and spatially-varying surface reflectance under
  unknown illumination.
\newblock {\em ACM Trans. Graph.}, 35(6), December 2016.

\bibitem{Xu:2016:MBS}
Z.~Xu, J.~B. Nielsen, J.~Yu, H.~W. Jensen, and R.~Ramamoorthi.
\newblock Minimal brdf sampling for two-shot near-field reflectance
  acquisition.
\newblock {\em ACM Trans. Graph.}, 35(6):188:1--188:12, Nov. 2016.

\bibitem{Ye:2018:SPS}
W.~Ye, X.~Li, Y.~Dong, P.~Peers, and X.~Tong.
\newblock Single photograph surface appearance modeling with self-augmented
  {CNNs} and inexact supervision.
\newblock {\em Computer Graphics Forum}, 37(7), Oct 2018.

\bibitem{Zhou:2016:SPS}
Z.~Zhou, G.~Chen, Y.~Dong, D.~Wipf, Y.~Yu, J.~Snyder, and X.~Tong.
\newblock Sparse-as-possible svbrdf acquisition.
\newblock {\em ACM Trans. Graph.}, 35, November 2016.

\end{thebibliography}
}

\end{document}